\documentclass[publicationstatus]{eptcs}
\usepackage{breakurl}             

\usepackage{amsfonts, amssymb, amscd, amsthm, amsmath, xspace, bbm}
\usepackage{IEEEtrantools}

\usepackage{algorithm}
\usepackage{subcaption}
\usepackage{algorithmic}

\usepackage{color}
\usepackage{graphicx}
\usepackage{epstopdf}

\usepackage{xfrac}

\newtheorem{theorem}{Theorem}
\newtheorem{lemma}{Lemma}

\newtheorem{remark}{Remark}

\newtheorem{proposition}{Proposition}

\newcommand{\etal}{{\it et al.}}
\newcommand{\ie}{{\it i.e.}}
\newcommand{\eg}{{\it e.g.}}

\newcommand{\GF}[1]{\mathbb{F}_{#1}} 

\newcommand{\hash}[1]{{\texttt{hash}\left(#1\right)}} 
\newcommand{\block}[1]{B_{#1}} 
\newcommand{\hblock}[1]{\hat{B}_{#1}} 
\newcommand{\header}[1]{H_{#1}} 
\newcommand{\payload}[1]{T_{#1}} 
\newcommand{\mroot}[1]{{\texttt{root}\left(#1\right)}} 
\newcommand{\size}[1]{{\texttt{size}\left(#1\right)}} 

\newcommand{\Enc}[1]{{\texttt{Enc}\left(#1\right)}} 
\newcommand{\Dec}[1]{{\texttt{Dec}\left(#1\right)}} 

\newcommand{\K}{K} 

\newcommand{\cblock}[2]{C^{(#1)}_{#2}} 

\newcommand{\cblockj}[1]{C_{#1}} 
\newcommand{\vj}[1]{v_{#1}} 
\newcommand{\tS}{\tilde{S}} 
\newcommand{\lt}{t} 
\newcommand{\Ls}{L_s} 

\newcommand{\bandwidth}[1]{\beta(#1)} 

\title{SeF: A Secure Fountain Architecture for Slashing Storage Costs in Blockchains}
\author{Swanand Kadhe, Jichan Chung, and Kannan Ramchandran\\
\institute{Department
of Electrical Engineering and Computer Sciences,\\ University of California,
Berkeley\\}
\email{\{swanand.kadhe, jichan3751, kannanr\}berkeley.edu}
}
\def\titlerunning{Coding for Blockchains}
\def\authorrunning{Kadhe \etal}

\begin{document}



\maketitle

\begin{abstract}
Full nodes, which synchronize the full blockchain history and independently validate all the blocks, form the backbone of any blockchain network by playing a vital role in ensuring security properties. On the other hand, a user running a full node needs to pay a heavy price in terms of storage costs. In particular, blockchain storage requirements are growing near-exponentially, easily outpacing Moore's law for storage devices. For instance, the Bitcoin blockchain size has grown over 215GB, in spite of its low throughput. The ledger size for a high throughput blockchain Ripple has already reached 8.4TB, and it is growing at an astonishing rate of 12GB per day! 

In this paper, we propose an architecture based on {\em fountain codes}, a class of erasure codes, that enables any full node to {\em encode} validated blocks into a small number of {\em coded blocks}, thereby reducing its storage costs by orders of magnitude.
In particular, our proposed {\em Secure Fountain (SeF)} architecture can achieve a {\em near optimal} trade-off between the storage savings per node and the {\em bootstrap cost} in terms of the number of (honest) storage-constrained nodes a new node needs to contact to recover the entire blockchain. 
A key technical innovation in SeF codes is to make fountain codes secure against adversarial nodes that can provide maliciously formed coded blocks. Our idea is to use the header-chain as a {\em side-information} to check whether a coded block is maliciously formed {\em while it is getting decoded}.
Further, the {\em rateless property} of fountain codes helps in achieving high decentralization and scalability. We evaluate the performance of the SeF architecture by performing experiments on the Bitcoin blockchain. 
Our experiments demonstrate that SeF codes tuned to achieve $1000\times$ storage savings enable full nodes to encode the 191GB Bitcoin blockchain into 195MB (on average). A new node can recover the blockchain from an arbitrary set of storage-constrained nodes as long as the set contains $\sim$1100 honest nodes (on average). Note that for a $1000\times$ storage savings, the fundamental bound on the number of honest nodes to contact is $1000$: we need about 10\% more in practice. 
More generally, SeF codes can achieve a continuum of trade-offs between storage savings and bootstrap cost to new nodes (number of honest storage-constrained nodes they have to contact) that is near-optimal. 
\end{abstract}


\section{Introduction}
\label{sec:intro}
Blockchains have played an instrumental role as the foundational technology for cryptocurrencies such as Bitcoin and Ethereum. Moreover, they have the potential to disruptively impact diverse fields such as the  Internet-of-Things~\cite{Blockchain:IoT:17}, medicine~\cite{Blockchain:medical:16}, healthcare~\cite{Blockchain:healthcare:16}, and supply-chains~\cite{Blockchain:supplychain:17} among others. This great potential of blockchains comes from their key differentiating properties of decentralization, security, trustlesssness, and scalability. (For simplicity, we refer to these properties as security properties.)

A blockchain network safeguards its security properties by relying on its nodes to independently validate every block added to the chain, store the entire blockchain history, and contribute in helping new nodes that want to join the network. Node with these functionalities---often called ``full nodes'' in the cryptocurrency parlance---form the backbone of any blockchain network, as they play a vital role in ensuring the security properties. More specifically, by independently verifying transactions and blocks without relying on any other node, full nodes contribute to the health of the network by safeguarding its security and trustlessness, and by helping to bootstrap\footnote{Henceforth, we will refer to ``bootstrap'' to mean ``providing a new node with the entire blockchain history to bring it up to speed''.} new nodes joining the network, they ensure  decentralization and scalability of the network. Indeed, full nodes are critical for any blockchain system's survival, and major cryptocurrencies typically recommend the users, which are running businesses, exchanges or block explorers, or participating in consensus (\ie, miners) to run full nodes to achieve complete security (see, \eg,~\cite{Bitcoin-wiki-full-node}).

\begin{figure}[!t]
    \centering
    \includegraphics[scale=0.32]{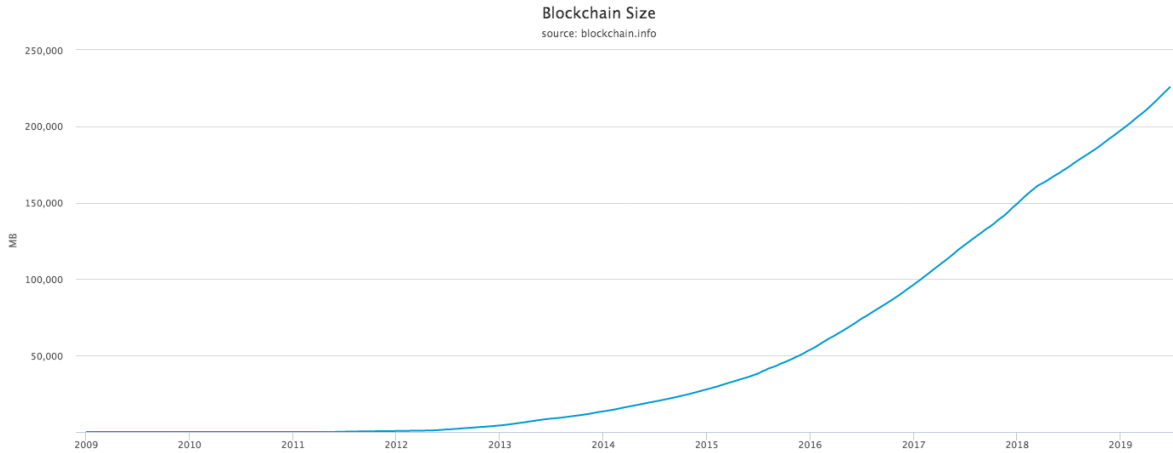}
    \caption{The total size of all block headers and transactions for the Bitcoin blockchain, not including database indexes (source:~\cite{Blockchain-info:18}).}
    \label{fig:bitcoin-size}
\end{figure}

On the other hand, a user running a full node needs to pay a heavy price in terms of storage and computation costs. In particular, blockchain storage requirements are growing near-exponentially, easily outpacing Moore's law for storage devices. To get a glimpse of the heavy costs required for storing the blockchain's historical data, consider the case of Bitcoin. In spite of its low throughput of just 4-7 transactions per second, the Bitcoin blockchain size has grown over $215GB$ as of April 2019~\cite{Blockchain-info:18} (see Fig.~\ref{fig:bitcoin-size}). 
In fact, {\it storage costs are going to be a pressing concern in the near future for high throughput blockchains like Ripple}. For instance, 
the Ripple (XRP) ledger size has already reached $8.4TB$, and it is growing at an astonishing rate of $12GB$ per day! (See~\cite{Ripple-size:18}.)  

\begin{figure}[!t]
    \centering
    \includegraphics[scale=0.35]{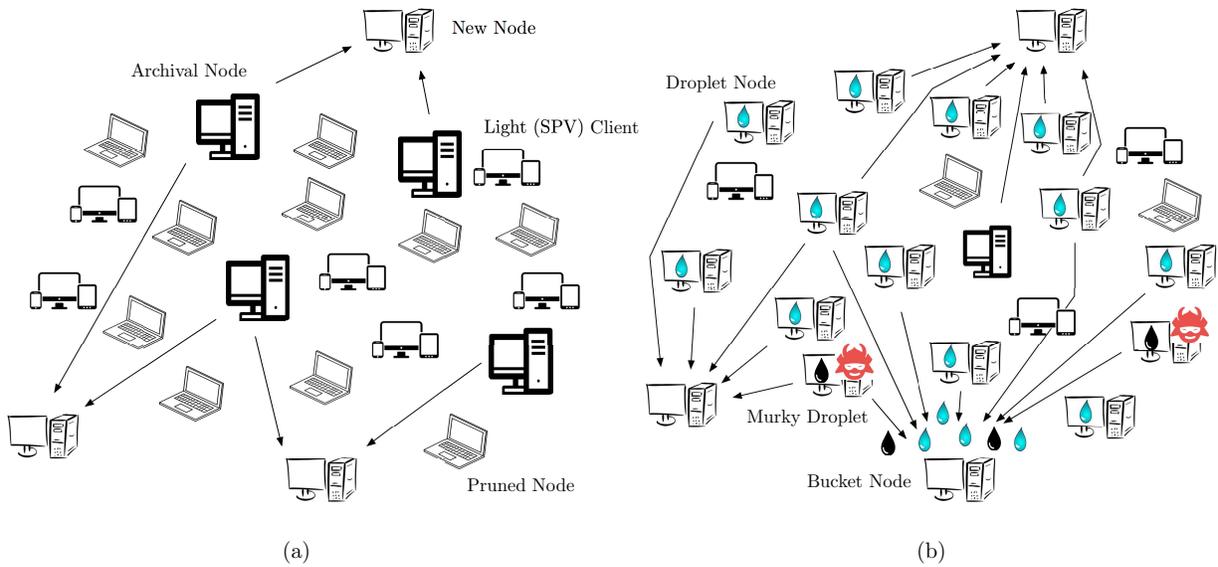}
    \caption{(a) {\it Current architecture for a blockchain network consists of archival nodes, pruned nodes, and light clients, out of which only archival nodes can help in bootstrapping a new node joining the network.} (b) {\it SeF architecture envisions a blockchain network mainly consisting of the proposed} {\bf droplet nodes} {\it that require low storage and computation resources.
    During bootstrap, a new node, called a} {\bf bucket node,} {\it collects sufficiently many droplets and recovers the blockchain even when some droplet nodes are adversarial, providing} {\bf murky (malicious) droplets}. {\it After validating the blockchain, a bucket node will perform encoding to turn itself into a droplet node. In this way, droplet nodes will slowly replace archival nodes.}}
    \label{fig:droplet-architecture}
\end{figure}

In current practice, there are two solutions for saving costs: (i) run a {\it light} or {\it thin} client, also known as simplified payment verification (SPV) client~\cite{Nakamoto:09,Bitcoin-wiki-spv,Ethereum-wiki-light-client}, or (ii) enable {\it block pruning}~\cite{Bitcoin-org-full-node}. 
Running a light client is the most economical way of saving costs. Light clients store only block headers, and do not validate transactions. However, light clients 
are known to be vulnerable to several security and privacy attacks (see, \eg,~\cite[Chapter 6]{Karame:16}). A pruned node stores only a budgeted number of most recent blocks, and deletes old blocks after they are validated. Though, unlike light clients, pruned nodes have strong security properties, they cannot contribute to scaling up the network in a secure and decentralized manner as they are unable to assist new full nodes. Indeed, if a large number of full nodes enable pruning, then new nodes will need to rely on a small number of {\it archival nodes} (\cite{Bitcoin-wiki-full-node}) that store the entire blockchain in order to bootstrap, greatly compromising the  decentralization requirement (see Fig.~\ref{fig:droplet-architecture} (a)).

Compelled by the essential role that full nodes play in ensuring the security properties and the heavy costs they incur, 
this paper presents {\it SeF, a Secure Fountain architecture founded on coding theory, that enables storage-constrained machines to act as full nodes without affecting the security properties of the blockchain.} 
Our main focus is on decreasing the cost of storing the blockchain's historical data, which is often much larger than that of storing its {\it state} (\eg, the state of Bitcoin, the so-called UTXO set, is around 3GB, as compared to its overall size of 215GB~\cite{Blockchain-info:18}). The key challenge in reducing the cost of storing blockchain's historical data is that it is required to bootstrap new nodes that join the network, and bootstrapping plays a key role in scaling up the security and decentralization capability of the network.


In particular, SeF must overcome the following challenges: 
\begin{itemize} 
\item {\it Security:} The protocol must ensure that the blockchain network can scale up in a {\it secure} manner even if a subset of storage-constrained full nodes are adversarial. Specifically, a new node should be able to recover the blockchain even if any (limited) subset of storage-constrained full nodes act adversarially and provide maliciously formed data to the new node. Moreover, the computational cost associated with recovering the blockchain must be small. 
\item {\it Decentralization:} The protocol must be {\it decentralized} allowing every full node to perform computations to reduce its storage space without relying on any other full node. 
\item {\it Bootstrap Cost:} The protocol must have limited {\it bootstrap cost} in terms of the number of storage-constrained full nodes that a new node needs to contact in order to recover the blockchain.
\end{itemize}

\begin{figure}[!t]
    \centering
    \begin{subfigure}[t]{0.5\textwidth}
        \centering
        \includegraphics[width=\textwidth]{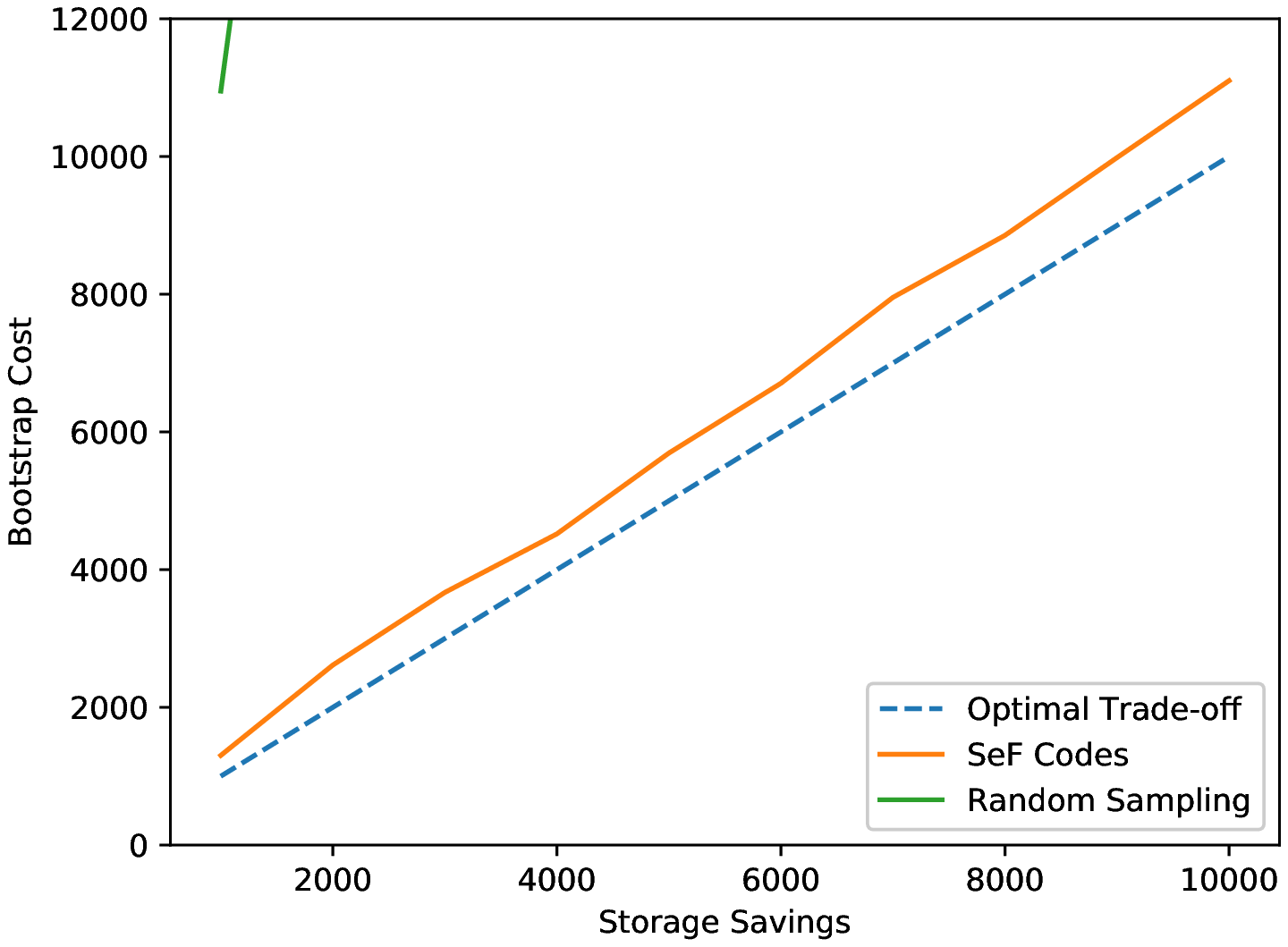}
        \caption{SeF Codes}
        \label{fig:bootstrap-cost-LT}
    \end{subfigure}%
    ~ 
    \begin{subfigure}[t]{0.5\textwidth}
        \centering
        \includegraphics[width=\textwidth]{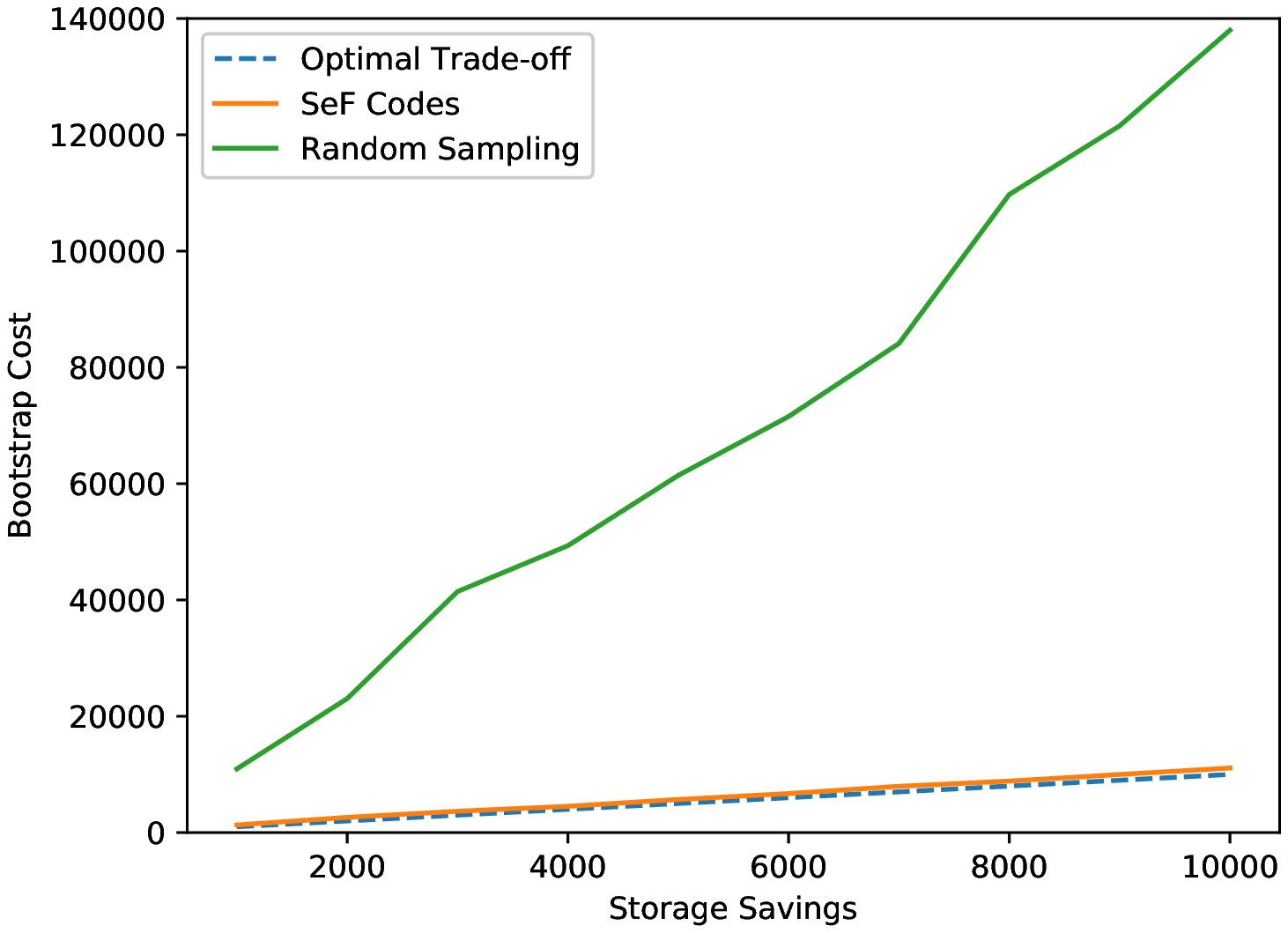}
        \caption{Random Sampling}
        \label{fig:bootstrap-cost-CC}
    \end{subfigure}
    \caption{{\it Theoretical and achievable trades-off between the bootstrap cost versus storage savings. We define {\bf bootstrap cost} as the number of storage-constrained full nodes (\ie, droplet nodes) that a new node needs to contact in order to recover the entire blockchain with high probability (we consider 99\% in the plots). The optimal (theoretical) trade-off is shown with a dashed line which depicts that for any scheme with $\gamma$-fold storage savings, the bootstraps cost is at least $\gamma$ (see Sec.~\ref{sec:analysis} for details.) Observe in plot (a) that our proposed SeF codes achieve a near-optimal trade-off. We also highlight the heavy bootstrap cost incurred by random sampling in plot (b).}}
    \label{fig:savings-vs-cost}
\end{figure}

In fact, there is a fundamental trade-off between the storage savings and the bootstrap cost as shown in Fig.~\ref{fig:savings-vs-cost} (dashed line).  Specifically, consider any scheme that enables full nodes to reduce their storage space to  $1/\gamma$ fraction of the blockchain size (for some positive real number $\gamma$). Then, a new node needs to contact at least $\lceil\gamma\rceil$ storage-constrained full nodes to recover the blockchain. This is simply because the total amount of data downloaded by a new node must be at least the size of the blockchain. As an example, consider a scenario in which every full node restricts its storage space to $1GB$. Then, a new node in the Bitcoin network will need to contact at least $215$ (honest) nodes to obtain the $215GB$ Bitcoin blockchain. Whereas, a new node in the Ripple network will need to contact at least 84,000 (honest) nodes to obtain the $8.4TB$ Ripple (XRP) ledger. In summary, the larger the storage savings per full node, the higher the bootstrap cost for a new node.

In  a centralized system, it is easy to keep the bootstrap cost to its minimum, for instance, by partitioning the blockchain across nodes. 
However, using na\"ive approaches to achieve decentralization can result in prohibitively high bootstrap cost. As an example, consider the following simple protocol for full nodes to cut down their storage space. For every $k$ blocks (say, $k = $ 10,000), a node stores a randomly selected block, independent of other nodes.\footnote{The Ripple blockchain uses a similar scheme called {\it history sharding} to save storage while contributing to preserving historical XRP Ledger data~\cite{Ripple-sharding:18}. In history sharding, the transaction history of the XRP Ledger is partitioned into segments, called shards. A server that has enabled history sharding acquires and stores randomly selected shards, where the number of stored shards depends on the budgeted storage space.} Each node thus achieves $k$-fold storage savings. However, it is not hard to show that, in this case, a new node requires to contact a lot more than $k$ nodes. In fact, obtaining the blockchain in this scheme is, in fact, identical to the classical ``coupon collector'' problem (see, \eg,~\cite[Chapter 3.6]{Motwani:95}), where there is a (multiplicative) logarithmic hit in the number of nodes needing to be contacted (see Fig.~\ref{fig:bootstrap-cost-CC}; green curve).
Therefore, it is of paramount importance to design decentralized schemes that achieve storage savings without incurring substantial bootstrap cost.

\subsection{SeF Codes In a Nutshell}
\label{sec:contributions}
SeF addresses the aforementioned challenges by enabling full nodes to {\it encode} validated blocks into a small number of {\it coded blocks}, thereby requiring significantly less storage space. The core of SeF is built up on a class of erasure codes called {\it fountain codes}~\cite{Byers:98,Luby:02} (see also~\cite{Mackay:05,Shokrollahi:now:11}). The {\it encoder} of a fountain code is a metaphorical fountain that takes as an input a set of blocks of fixed size and produces a potentially endless supply of {\it water drops} (\ie, {\it coded blocks}). Anyone who wishes to recover the original blocks holds a {\it bucket} under the fountain and collects drops until the number of drops in the bucket is slightly larger than the number of original blocks. They can then {\it decode} the original blocks from the collected drops. 


A key technical innovation in SeF codes is to make fountain codes secure against adversarial nodes (hence, the name {\it Secure Fountain} codes).\footnote{Fountain codes have originally been designed to cater to random erasures, and cannot be directly used to correct adversarial errors. See Sec.~\ref{sec:related-work} for details.} Fountain codes admit a computationally efficient decoding process, called a {\it peeling decoder}~\cite{Luby:02} (also known as a {\it belief propagation}; see, \eg,~\cite{Richardson:08}). A peeling decoder is an iterative decoder that decodes one block in each iteration and {\it peels off} (removes) its contribution from the remaining coded blocks. SeF codes introduce error-resiliency in the peeling process by enabling the decoder to identify maliciously formed encoded blocks. 
In essence, the idea is to use the header-chain as a side-information and leverage Merkle roots stored in block-headers to check whether a coded block is maliciously formed {\it while it is getting decoded}. Indeed, the peeling decoder turns out to be crucial in identifying maliciously formed droplets, and thus, achieving high security.

Fountain codes are {\it rateless} in the sense that it is possible to produce a potentially limitless number of drops (coded blocks) from a fixed number of blocks.\footnote{The term {\it rateless} comes from the contrasting nature of fountain codes as compared to classical erasure codes (such as Reed-Solomon codes; see~\cite{MacWilliams-Slone:78}), in which a set of blocks of fixed size is encoded into a larger set of coded blocks of that is also of fixed side. The ratio of the number of coded blocks to the number of original blocks is called the rate of the code.} SeF codes inherit the rateless property from fountain codes, which allows each node to produce coded blocks without relying on other nodes. Therefore, SeF codes are {\it decentralized}, making {\it every node} useful for bootstrapping a new node. 

Our proposed {\it SeF codes} create a blockchain network consisting of full nodes with low storage resources, referred to as {\it droplet nodes} (see Fig.~\ref{fig:droplet-architecture} (b)). Every droplet node independently {\it encodes} validated blocks into a small number of {\it droplets} (\ie, coded blocks) using a fountain code, thereby requiring significantly less storage space. To recover the blockchain during bootstrap, a new node acts like a {\it bucket}, and collects sufficiently many droplets by contacting any arbitrary subset of droplet nodes. (Hence, the terms droplets and droplet nodes, as any droplet is as useful as the other!) Even if a fraction of droplet nodes act adversarially and provide maliciously formed droplets (called {\it murky droplets}), our proposed decoding can identify such murky droplets and delete them. Finally, the new (bucket) node turns itself into a droplet node by validating blocks and encoding the blockchain into droplets, and the process continues.

SeF codes can achieve a near optimum trade-off between the storage savings and the bootstrap cost. In particular, SeF codes allow the network to tune the storage savings as a parameter, depending upon how much bootstrap cost new nodes can tolerate. When  SeF codes are tuned to achieve $k$-fold storage savings, a new node is guaranteed to recover the blockchain with probability $(1-\delta)$ by contacting $k + O(\sqrt{k}\ln^2(k/\delta))$ honest nodes. In fact, our experiments show much better results as shown in Fig.~\ref{fig:savings-vs-cost} (orange curve). 

\subsection{Related Work}
\label{sec:related-work}
Bitcoin allows full nodes to reduce their storage costs by enabling block pruning~\cite{Bitcoin-org-full-node}. However, pruned nodes cannot help new nodes to join the network and do not contribute in preserving the historical blockchain data. Ethereum uses state tree pruning~\cite{Ethereum-state-pruning:15} to reduce storage overhead, however, full nodes typically store the entire blockchain. A recent proposal~\cite{Ethereum-chain-pruning:18} for pruning the Ethereum blockchain discusses several ways of scaling storage requirements, such as offloading the historical blockchain data to decentralized archives such as IPFS, Swarm, or BitTorrent. On the other hand, SeF codes enable full nodes to reduce their storage costs in such a way that they can still contribute in bootstrapping new nodes and preserving the blockchain history. 

Ripple uses a {\it random sampling} scheme, referred to as {\it history sharding}, for enabling servers to reduce their storage in such a way that the ledger history is still preserved by the network~\cite{Ripple-sharding:18}. In particular, the transaction history of the XRP Ledger is partitioned into segments, called shards. A server that has enabled history sharding acquires and stores randomly selected shards. As we discuss in Sec.~\ref{sec:analysis-others}, random sampling results in significant bootstrap cost, whereas SeF codes achieve near-optimal bootstrap cost. 

It is worth noting that, in a conventional blockchain network, every full node stores the entire history of the blockchain. From the perspective of storage, such a network can be viewed as a distributed storage system with replication. As erasure codes are known to be greatly successful in reducing storage costs in distributed storage systems without reducing reliability~\cite{Dimakis:surevey:11,Plank:surevey:13,Huang:Azure:12}, it is natural to consider erasure codes to reduce storage costs in blockchains. This idea is considered in~\cite{Perard:storage:18,Dai:storage:18,Raman:storage:17,Polyshard:18}. 

In particular, references~\cite{Perard:storage:18,Dai:storage:18} propose {\it low-storage nodes} which split every block into small, fixed-sized fragments, and store only {\it coded fragments}. These coded fragments are obtained by linearly combining the block fragments with random coefficients. The main limitation of these works is that they only consider the the case when nodes can leave the network or can be unreachable; they do not consider adversarial nodes that can provide maliciously formed coded fragments. 

In~\cite{Raman:storage:17}, the authors consider the problem of storing a blockchain with confidentiality and reduced storage. They propose to first dynamically partition the network into zones. Then each block is encrypted with a key specific to a zone and the encrypted block is distributed across the nodes in a zone using a distributed storage code, such as~\cite{Dimakis:surevey:11,Huang:Azure:12}.

In~\cite{Polyshard:18}, the authors consider a sharded blockchain, and propose to compute a {\it coded shard} by linearly combining uncoded shards. In particular, Reed-Solomon codes (see, \eg,~\cite{MacWilliams-Slone:78}) are used to generate the coded shards. With Reed-Solomon codes, it is possible to recover the original data in the presence of (a limited number of) adversarial nodes providing malicious data~\cite{MacWilliams-Slone:78}. 

All these coding schemes -- random linear codes, distributed storage codes, and Reed-Solomon codes -- need to operate over a sufficiently large finite field, and incur high computational complexity for decoding. On the other hand, SeF codes are based on fountain codes, especially LT codes, which are substantially better in terms of computational cost (see Sec.~\ref{sec:analysis-others}).

It is important to note that fountain codes have been designed to handle (random) erasures. While it is possible to decode from random errors (see, \eg,~\cite{Etesami:06,Luby:verfication:05,Karp:verification:05}), adversarial errors can be difficult to deal with.\footnote{Techniques proposed to handle adversarial errors such as~\cite{Juels:falcon:15} require shared secret between the encoder and the decoder. This is not possible in a blockchain network since nodes are supposed to encode the blockchain in a decentralized manner.} In general, iterative decoding algorithm for fountain codes will readily propagate (and amplify) any error in the received data into the recovered data. This is because fountain codes do not provide any mechanism for checking the integrity of the decoded data. The key observation of this paper is that the Merkle root of a block together with the header-chain structure of a blockchain enables one to check the integrity of the decoded blocks. 

\section{System Overview}
\label{sec:overview}

\subsection{Blockchain Model}
\label{sec:blockchain}

\begin{figure}[!t]
    \centering
    \includegraphics[scale=0.450]{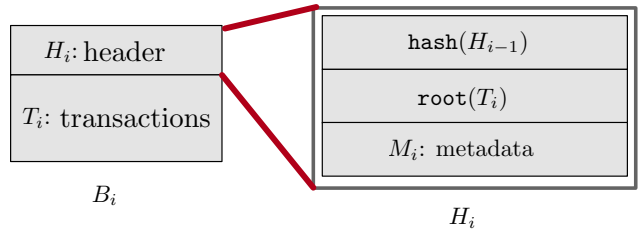}
    \caption{{\it Structure of a block and its header.}}
    \label{fig:block-structure}
\end{figure}

 A blockchain is simply a sequence of blocks chained together using cryptographic hashes. Each block contains a list of transactions and a header. In particular, we consider the following generalized structure of a block (see Fig.~\ref{fig:block-structure}).
\begin{itemize}
    \item Let $\hash{\cdot}$ denote a cryptographic hash function (such as SHA-256).
    \item Let $\mroot{T}$ denote the Merkle (tree) root\footnote{A Merkle tree is a balanced binary tree where the value of each non-leaf node is the hash of its children~\cite{Merkle:80}.} of a list of items $T$.
    \item The $i$-th block $\block{i}$ in the blockchain is denoted as $\block{i} = \{\header{i},\payload{i}\}$, where the payload $\payload{i}$ is a list of transactions, and the header $\header{i} = \{\mroot{\payload{i}}, \hash{\header{i-1}},M_i\}$, where $M_i$ denotes metadata such as timestamp and consensus related information (the exact contents of the metadeta are not relevant here). We set $\hash{\header{-1}} = 0$ as a convention.
\end{itemize}
For simplicity, we assume that each block is of size $L$ bits.\footnote{We discuss how to handle variable block sizes in Sec.~\ref{sec:block-sizes}.} 
Further, we assume that the first $L_h$ $(< L)$ bits of the block correspond to its header, whereas the remaining $L - L_h$ bits correspond its payload.

\noindent{\bf Mining and Consensus:} Blocks are created and appended to the blockchain via a {\it mining process}, where the participating nodes, known as {\it miners}, compete to become the next block proposer. A typical way to compete is by solving a computationally-intensive puzzle, known as {\it proof-of-work}, with sufficient difficulty.
A blockchain network uses a consensus algorithm to determine which chain should be selected in case there is a fork. For the clarity of exposition, we focus our attention to the proof-of-work based Nakamoto consensus~\cite{Nakamoto:09} in the paper.\footnote{We discuss how the proposed coding scheme can be applied to other types of consensus algorithms such as proof-of-stake in Sec.~\ref{sec:discussion}.} 
In the Nakamoto consensus, the chain with the most accumulated work (referred to as the {\it longest chain}) is selected in the event of a fork. In addition, there are protocol rules to determine the validity of transactions and blocks. 

\noindent{\bf Full Nodes:} A typical node in a blockchain network, referred to as a full node, stores a copy of the entire blockchain, and validates new blocks as well as transactions. 
Whenever a new full node joins the network, it first needs to synchronize to the current {\it state} (\eg, account balances) by downloading and validating the blockchain until that time.\footnote{This is typically referred to as {\it full synchronization}. A blockchain may offer other faster ways of synchronization (\eg, fast synchronization in Ethereum). However, the full synchronization is the most secure way to join a blockchain network~\cite{Ethereum-fast-sync}.} A typical full node stores the entire blockchain to help bootstrap new nodes, and for preserving the history.


\subsection{Threat Model and Problem Formulation}
\label{sec:model}
We are interested in designing protocols that significantly reduce the storage costs at full nodes. 
There are two key components associated with blockchain storage costs: (a) The cost of storing the current state that is necessary for validating the content getting added. For example, the state can be all currently spendable transactions (\eg, Bitcoin) or all current account balances (\eg, Ethereum). This essentially is the information necessary for full nodes to perform transaction validation. (b) The cost of storing the blockchain's historical data. This is necessary to bootstrap new nodes that join the network, and is often much larger than the state. For example, the size of the Bitcoin state is around 3GB, as compared to its overall size of 215GB~\cite{Blockchain-info:18}.

In this work, we focus our attention to reducing storage costs associated with storing the blockchain's historical data. Our goal is to design a protocol that enables a full node to reduce its storage space in such a way that the node is still able to help in bootstrapping a new node. We refer to a node with reduced storage space as a {\it droplet node}, and a new node joining the system as a {\it bucket node}. 

\vspace{2mm}
\noindent {\bf Threat Model:} 
We consider a Byzantine adversary that can control an arbitrary subset of droplet nodes. These malicious droplet nodes may collude with each other and can deviate from the protocol in any arbitrary manner, \eg, by storing/sending arbitrary data to a bucket node, or staying silent. The remaining nodes are honest and faithfully follow the protocol. We assume that the  adversary is oblivious, \ie, it does not observe the storage contents of droplet nodes before choosing which nodes to control.
{\it Our goal is to design protocols that allow a bucket node to reconstruct the blockchain as long as a {\it small} number of droplet nodes are honest.} We measure the security performance of a coding scheme by the minimum number of honest droplet nodes that are sufficient to recover the blockchain with overwhelming probability. 

Our proposed scheme assumes that a bucket node can first obtain the honest (correct) header-chain. Towards this end, we assume that the majority of the consensus (\ie, block producing nodes or miners) is honest. Further, we assume that the adversary is computationally bounded, and cannot construct a longer chain than the one constructed by the honest consensus.

\vspace{2mm}
\noindent {\bf Problem Formulation:} Let $\lt$ denote the current height of the (longest) blockchain, and let $B = \{\block{1},\block{2},\ldots,\block{\lt}\}$. 
For an arbitrary subset of blocks $B' \subseteq B$, let $\size{B'}$ denote the size of $B'$ in bits. Let $\gamma$ be a positive real number greater than $1$.
Our goal is to design a pair of encoding and decoding schemes $(\texttt{Enc},\texttt{Dec})$, referred to as a coding scheme, for a target storage savings of $\gamma$ with the following properties:
\begin{itemize}
    \item[1.] $\texttt{Enc}$ is a (randomized) {\it encoding scheme} that enables a full node to reduce its storage space by a factor of $\gamma$. In particular, node $j$ computes and stores $\cblockj{j} = \Enc{B,j}$ such that $\size{B}/\size{\cblockj{j}} = \gamma$. We refer to the {\it coded blocks} $\cblockj{j}$ as {\it droplets}, and any node storing droplets as a {\it droplet node}.
    
    As an example, using the proposed SeF codes, a droplet node can encode $191.48GB$ of the Bitcoin blockchain into $195.6MB$ droplets.
    
    \item[2.] $\texttt{Dec}$ is a {\it decoding scheme} that allows a {\it bucket node} -- a new node joining the network -- to recover the blockchain $B$ from an arbitrary set of droplet nodes that contains a sufficient number of honest droplet nodes. 
    Specifically, there exist positive integers $K$, $n$ $(\geq K)$ such that, for an arbitrary set of droplet nodes $\{j_1,j_2,\cdots,j_n\}$ that contains at least $K$ honest ones,  $\Dec{\cblockj{j_1},\cblockj{j_2},\cdots,\cblockj{j_n}} = B$ with overwhelming probability. 
     
    As an example, in our proposed SeF scheme targeted at achieving $1000\times$ storage savings, a bucket node can recover the blockchain with high probability from $K \approx 1100$ honest droplet nodes.
\end{itemize}
In general, our goal is to design coding schemes that achieve 
small $K$ for a given storage savings $\gamma$.  


\vspace{2mm}
\noindent {\bf Performance Metrics:}
We measure the performance of a coding scheme using the following metrics.

\begin{itemize}
    \item[1.] 
    {\it Storage Savings} of a node is the ratio of the total blockchain size to the size of the droplets it stores. 
    
    \item[2.] 
    {\it Bootstrap Cost} of a coding scheme is measured by the minimum number of honest droplet nodes that a bucket node needs to contact in order to ensure that the blockchain can be recovered with overwhelming probability. Note that the bootstrap cost of a coding scheme reflects its {\it security performance}. This is because the bootstrap cost can be considered as the minimum number of honest droplet nodes that the system must contain to guarantee, with high probability, that the historical blockchain data is preserved. 
    The smaller the bootstrap cost of a coding scheme, the better the security performance of the system using the scheme. 

     \item[3.] 
    {\it Bandwidth Overhead} is the overhead in terms of the amount of data that a bucket node needs to download for recovering the blockchain with high probability. 
     
     \item[4.] 
    {\it Computation Cost} of a coding scheme is measured in terms of the number of arithmetic operations associated with the encoder $\texttt{Enc}$ and the decoder $\texttt{Dec}$.
     
\end{itemize}

\noindent {\bf Design Objectives:}  As mentioned in the introduction, it is straightforward to show that there is a fundamental trade-off between the storage savings and the bootstrap cost (see Sec.~\ref{sec:analysis} for details). Our main goal is to design protocols that can achieve a near-optimal trade-off between the storage savings and the bootstrap cost. Further, we want the protocols to have small bandwidth overhead and computational cost.
    In addition, we are interested in designing encoding schemes that are {\it decentralized}. Specifically, a droplet node should be able to generate its droplets without knowing what any other node in the system is storing.
    

\section{Secure Fountain Architecture}
\label{sec:SeF-main}

\subsection{Generic Framework}
\label{sec:framework}

We begin with a generic framework for a coding scheme, which enables a node to {\it code} across blocks and save its storage space by storing only a small number of {\it coded blocks}. Recall that we refer to the coded blocks as {\it droplets}, the nodes storing coded blocks as {\it droplet nodes}, and any new node joining the system as a {\it bucket node}.

\vspace{2mm}
\noindent {\bf (a) Encoding:} We propose to compute droplets in epochs, where an epoch is defined as the time required for the blockchain to grow by $k$ blocks (\eg, $k = 10000$). In the current epoch, when the blockchain grows by $k$ blocks, the sub-chain of length $k$ is {\it encoded}
into $s$ {\it droplets} \ie, {\it coded blocks} (\eg, $s = 10)$. Then, the encoding process continues to the next epoch. To handle blockchain reorganizations due to potential forks, the most recent $\tau$ blocks are excluded from encoding and are stored in an uncoded format (\eg, $\tau = 550$).\footnote{In the Bitcoin blockchain, a pruned node is required to store at least $550$ blocks so that it can handle forks.} In addition, each node stores the header-chain for the original blockchain.

\begin{figure}
    \centering
    \includegraphics[scale=0.5]{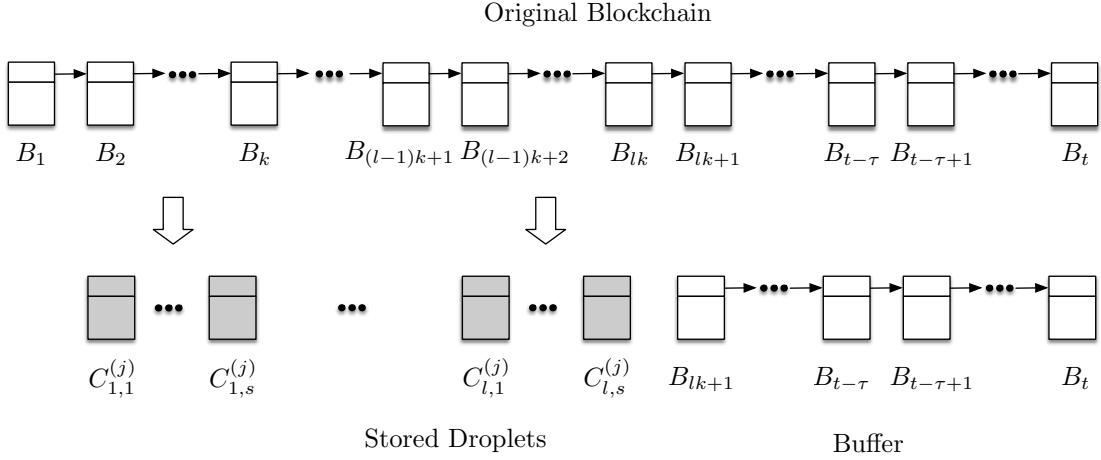}
    \caption{{\it Encoding happens in} {\bf epochs}. {\it An epoch is defined as the time required for the blockchain to grow by $k$ blocks. In the current epoch, when the blockchain grows by $k$ blocks, the sub-chain of length $k$ is encoded into $s$ droplets. Then, the encoding process continues to the next epoch.} 
    {\it For example, for $k = 10000$ and $s = 10$, each droplet node} {\bf reduces its storage cost by a factor of $\mathbf{1/1000}$}. {\it This means a node can encode the Bitcoin blockchain of size 190GB into little over 190 MB.}
    }
    \label{fig:encoding}
\end{figure}

More specifically, the first epoch starts from the $(\tau+1)$-th block. When the blockchain grows up to block $\block{k+\tau}$, a node encodes the blocks $\block{1}, \block{2},\ldots, \block{k}$ into $s$ droplets. The node then deletes the $k$ original blocks, and stores only the $s$ droplets for the first epoch. The process then continues into the next epoch. 
Let us denote the $s$ droplets stored by node $j$ in epoch $l$ as $\cblock{j}{l,1},\cblock{j}{l,2},\ldots,\cblock{j}{l,s}$. 
See Fig.~\ref{fig:encoding} for a schematic representation.


\noindent {\bf (b) Decoding:} Consider {\it a bucket node} joining the system when the height of the blockchain is $t$. 
Let $e = \lfloor (t - \tau) / k \rfloor$. 
The bucket node first contacts an arbitrary subset of $n$ droplet nodes (of sufficient size), and collects (downloads) their droplets for epochs $1\leq l\leq e$. The bucket node also downloads the uncoded blocks (from $\block{ek+1}$ onward) from one or more of the $n$ droplet nodes.

The encoding should be performed in such a way that the bucket node can recover the blockchain from the collected droplets.
In particular, let us denote the $n$ droplet nodes that are contacted as $\{j_1,j_2,\ldots,j_n\}$. Then, for every epoch $1\leq l\leq e$, the bucket node should be able to {\it decode} the sub-chain $\{\block{(l-1)k+1}$, $\block{(l-1)k+2}$, $\ldots$, $\block{lk}\}$ from the $ns$  droplets $\{\cblock{j_i}{l,p} : 1\leq i\leq n, 1\leq p\leq s\}$. 


\subsection{Secure Fountain (SeF) Codes}
\label{sec:SeF-codes}

We propose to perform the encoding using a Luby Transform (LT) code~\cite{Luby:02}. At the core of LT codes lies the concept of a {\it fountain code}~\cite{Byers:98}. A fountain code takes as an input a vector of $k$ input symbols, and produces a potentially limitless stream of output symbols.\footnote{Here, a symbol refers to a sequence of bits, and all symbols are assumed to be of the same size. Note that a block can be considered as a symbol.} The main property that is required of a fountain code is that it should be possible to recover the $k$ input symbols from any set of $K$ $(\geq k)$ output symbols with high probability. The parameter $K$ is desired to be very close to $k$. 

LT codes admit a computationally efficient decoding procedure known as {\it peeling decoder} (also known as a {\it belief propagation})~\cite{Richardson:08}. However, the peeling decoder is designed to decode in the presence of erasures and it cannot handle maliciously produced output symbols. Our key observation is that the peeling process can be exploited to introduce resiliency against maliciously formed blocks by using the header-chain as a side-information and leveraging Merkle roots stored in block-headers.
We refer to LT codes with the error-resilient peeling decoder as {\it Secure Fountain (SeF) codes}. 



\subsubsection{Encoder of a Luby Transform (LT) Code}
\label{sec:encoding}

In every epoch, a droplet node computes a droplet as follows. The node first {\it flips its private coins} to generate a random number $d$ between $1$ and $k$. Then, it selects $d$ out of $k$ blocks uniformly at random. Finally, it computes a bit-wise XOR of these $d$ blocks to obtain a droplet. The node stores the droplet along with the indices of the $d$ blocks used to obtain the droplet. This process is repeated to compute each of the $s$ droplets. 

\begin{figure*}[!t]
    \centering
    \includegraphics[scale=0.5]{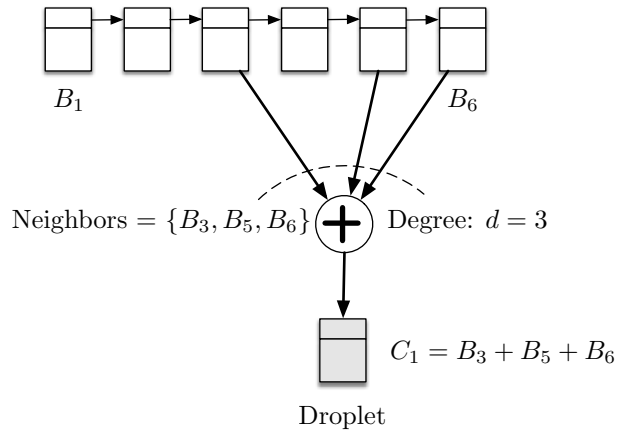}
    \caption{{\it An example for the LT code encoder. To generate a droplet in an epoch, a node first randomly samples a degree $d \in \{1,2,\ldots,k\}$ using the degree distribution (see Sec.~\ref{sec:degree-distribution}). Then, it chooses, uniformly at random, $d$ blocks from the epoch, and computes a bit-wise XOR of these blocks. These $d$ blocks are called the neighbors of the droplet.}}
    \label{fig:encoding-droplet}
\end{figure*}

In the terminology of LT codes, the number $d$ is refereed to as the {\it degree} of a droplet and the blocks used to compute a droplet are referred to as its {\it neighbors}. This terminology stems from considering a bipartite graph, with $k$ original blocks as left vertices and $s$ droplets as right vertices, in which there is an edge joining a block to a droplet if the block is used to compute the droplet. Further, the probability distribution on $\{1,2,\ldots,k\}$ used to sample degrees is referred to as the {\it degree distribution}.\footnote{We will describe the degree distribution used in SeF codes in Sec.~\ref{sec:degree-distribution}.} See Fig.~\ref{fig:encoding-droplet} for an example.

Now, we describe the encoding process formally. For simplicity, we focus our attention to the first epoch in the following. The encoding procedure is the same for all subsequent epochs.
A droplet node computes its $j$-th droplet $\cblockj{j}$, $1\leq j\leq s$, independent of the other droplets, as follows.
\begin{enumerate}
    \item Randomly choose the {\it degree} $d$ of the droplet from the degree distribution $\mu(\cdot)$. 
    
    \item Choose, uniformly at random, $d$ distinct blocks from the $k$ blocks, and set the droplet $\cblockj{j}$ as 
    the bit-wise XOR of these $d$ blocks. (These $d$ blocks are called {\it neighbors} of $\cblockj{j}$.) 
    
    Denote $\cblockj{j} = \{\header{j},\payload{j}\}$, where $\header{j}$ are the first $L_h$ bits of $\cblockj{j}$, referred as its header, and $\payload{j}$ are the remaining $L-L_h$ bits of $\cblockj{j}$, referred as its payload.\footnote{Note that the header and payload of a coded block may not have any semantic meaning.}
    
    \item Store $\cblockj{j}$ along with a length-$k$ binary vector $\vj{j}$ computed as follows: if the $m$-th block $\block{m}$ is among the $d$ blocks chosen to compute $\cblockj{j}$ then the $m$-th entry of $\vj{j}$ is $1$, else it is $0$.
\end{enumerate}

In addition to $s$ droplets, each droplet node stores the header-chain $H$ for the original blockchain.
As we will see, vector $\vj{j}$ and header-chain $H$ are required in the decoding process. In particular, $\vj{j}$ will be used to identify which original blocks are combined to generate $\cblockj{j}$, while the header-chain will enable the decoder to identify maliciously formed droplets. 
 
\begin{remark}
\label{rem:header-of-coded-block}
There are other, potentially more efficient, ways to convey which original blocks are combined to compute a droplet $\cblockj{j}$ than storing the length-$k$ binary vector. For instance, it is possible to store a seed using which a pseudo-random generator can produce the binary vector $\vj{j}$. We refer the reader to~\cite{Luby:02,Mackay:05} for more details. Since storing $\vj{j}$ takes much smaller size (\eg, 1250 bytes for $k = 10000$) as compared to typical block size (\eg, 1MB), we do not consider other methods. 
\end{remark}

\subsubsection{Adversarial Behavior Against SeF Codes}
\label{sec:adversarial-behavior}
We outline how an adversarial droplet node can behave in the SeF architecture. In addition to staying silent when contacted by a bucket node, an adversarial droplet node can act maliciously in the following two ways:
\begin{itemize}
    \item Store arbitrary values for $\cblockj{l}$, $\vj{l}$, and $H$. In particular, for a specific epoch, let $\mathbf{B}$ be a $k \times L$ binary matrix, in which the $i$-th row corresponds to the $i$-th block in the epoch. Then, for an honest node $j$, $\vj{j}$ and $\cblockj{j}$ are such that $\cblockj{j} = \vj{j}\mathbf{B}$. On the other hand, an adversarial node $l$ can store any values for $\cblockj{l}$ and $\vj{l}$ such that $\cblockj{l} \ne \vj{l}\mathbf{B}$. We refer to such a droplet as a {\it murky droplet}.
    \item Arbitrarily choose degree $d$, and arbitrarily choose $d$ blocks to compute a droplet. Store the coded block $\cblockj{j}$ and the vector $\vj{j}$ correctly. We refer to such a droplet as an {\it opaque droplet}.
    This attack is essentially targeted at increasing the probability of decoding failure. 
\end{itemize}
We refer to the droplets computed by honest nodes as {\it clear droplets}.

\subsubsection{Error-Resilient Peeling Decoder}
\label{sec:decoding}
Consider a bucket node that is interested in recovering the blockchain $B$. 
It contacts an arbitrary subset of $n$ $(n\geq k)$ droplet nodes, and downloads the stored data. This includes  droplets $\cblockj{j}$'s and vectors $\vj{j}$'s.
Without loss of generality, let us (arbitrarily) label the downloaded droplets as $\cblockj{1},\cblockj{2},\ldots,\cblockj{ns}$. Note that, since a coded droplet does not have any semantic meaning, the bucket node cannot differentiate between the clear, murky, and opaque droplets within the downloaded ones.

We assume that the bucket node has access to the honest header-chain. 
Note that this is can simply be done by contacting several droplet nodes, and obtaining the longest valid header-chain. We discuss the details in Sec.~\ref{sec:header-chain}.
Then, the node leverages this header-chain to perform error-resilient peeling decoding for an LT code, described as follows. 

The decoding proceeds in iterations. In each iteration the algorithm decodes (at most) one block until all the blocks are decoded, otherwise the decoder declares failure. We first describe the algorithm and then present a toy example. Let $\{\header{1}, \header{2},\ldots,\header{k}\}$ denote the first $k$ headers from the honest header-chain.

\begin{figure}[!h]
    \centering
    \includegraphics[scale=0.7]{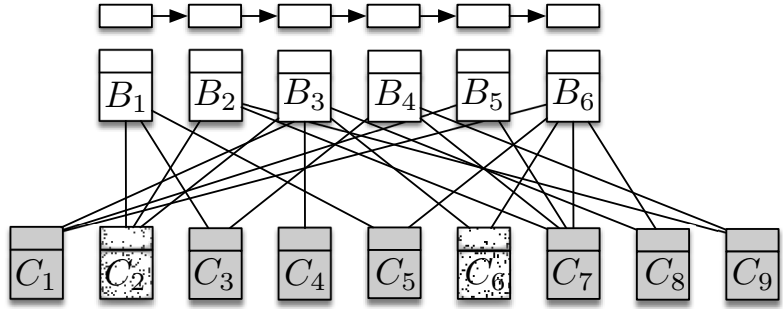}
    \caption{Toy example for the error-resilient peeling decoder for $k=6$ blocks and $ns = 9$ droplets. The bipartite graph $G$ at the initialization Step (1).}
    \label{fig:peeler-1}
\end{figure}

\begin{enumerate}
    
    
    \item {\it Initialization:} Form a bipartite graph $G$ with the $k$ original blocks as left vertices and the $ns$ droplets as right vertices. There is an edge connecting a droplet $\cblockj{j}$ to an original block $\block{m}$ if $\block{m}$ is used in computing $\cblockj{j}$. (Recall that this can be identified using $\vj{j}$. See Fig.~\ref{fig:peeler-1} for a toy example.)
    
    Set $\hblock{m} = \texttt{NULL}$ for $m = 1,2,\ldots,k$, where $\texttt{NULL}$ denotes null value.
    
    Set iteration number $i = 1$ and $G^{i-1} = G$.
    
    
    \item Find a droplet $\cblockj{l}$ that is connected to {\it exactly one} block $\block{m}$ in $G^{i-1}$. (Such a droplet is called a {\it singleton}.) 
    
    If there is no singleton, the decoding halts and declares failure. 
    
        
        
        
    
    \item Let $\header{l}$ and $\payload{l}$ be the header and payload of $\cblockj{l}$, respectively. 
    \begin{enumerate}
        \item Compute the Merkle root of $\payload{l}$, denoted as $\mroot{\payload{l}}$. If $\header{l}$ matches with the header $\header{m}$ in the header-chain $H$ and if $\mroot{\payload{l}}$ matches with the Merkle root stored in $\header{m}$, then set $\hblock{m} = \cblockj{l}$. (In this case, the droplet $\cblockj{l}$ is said to be {\it accepted}, and the $m$-th block is said to be decoded to $\hblock{m}$.) 
        
        \item Otherwise, delete $\cblockj{l}$ together with all its incoming edges from $G^{i-1}$ to obtain $G^{i}$. (In this case, the droplet $\cblockj{l}$ is said to be {\it rejected}.) 
        
        Increment $i$ by $1$.
        
        Go to Step (2).
    \end{enumerate}
    
    \item For all droplets $\cblockj{l'}$ that are connected to $\block{m}$ in $G^{i-1}$, set $\cblockj{l'} \leftarrow \cblockj{l'} \oplus \hblock{m}$. (Here, $\oplus$ denotes the bit-wise XOR.)
    
    \item Remove all the edges connected to the block $\block{m}$ from $G^{i-1}$ to obtain $G^{i}$. 
    
    \item Increment $i$ by $1$.
    
    \item If all the original blocks are not yet decoded, go to Step (2).
\end{enumerate}
Note that Step (3) differentiates the error-resilient peeling decoder from the classical peeling decoder for an LT code~\cite{Luby:02}.
More specifically, the classical peeling decoder always {\it accepts} a singleton, whereas the error-resilient peeling decoder may {\it reject} a singleton if its header and/or Merkle root does not match with the one stored in the header-chain. 

Note that at the initialization phase, it is not possible to determine whether a droplet is clear or murky if the droplet is not a singleton. However, when a droplet becomes a singleton, verifying whether its header matches with the corresponding one in the header-chain and  whether the Merkle root of its payload matches with the one stored in the corresponding header in the header-chain provides a mechanism for checking the integrity of the droplet.  This signifies the importance of singletons and underlines how crucial the peeling process is for achieving error-resiliency.

Next, we present a toy example for the decoder.

\noindent {\bf Toy Example:} We describe the decoder algorithm on the example shown in Fig.~\ref{fig:peeler-1}. We consider the epoch size of $k = 6$ blocks, and suppose that the bucket node has collected $9$ droplets, denoted as $\cblockj{1},\cblockj{2},\ldots,\cblockj{9}$. The corresponding bipartite graph $G$ is shown in Fig.~\ref{fig:peeler-1}. Suppose droplets $\cblockj{2}$ and $\cblockj{6}$ are murky. Note that the decoder does not know this at the beginning of the decoding. We assume that the bucket node has access to the honest header-chain, and denote its first $k = 6$ headers as $\{H_1,H_2,\ldots,H_6\}$. 

\begin{figure}[!h]
    \centering
    \includegraphics[scale=0.7]{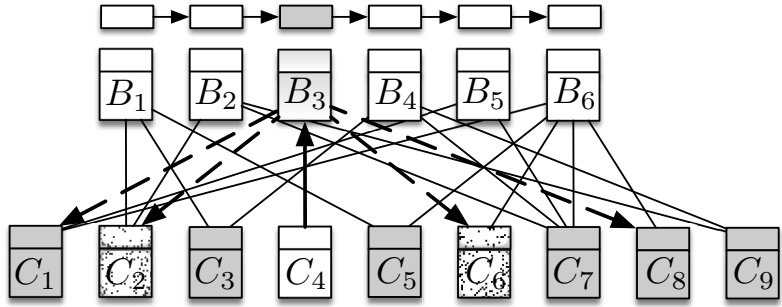}
    \caption{Iteration 1 with the bipartite graph $G^0$: the decoder accepts $\cblockj{4}$ and decodes $\block{3}$.}
    \label{fig:peeler-2}
\end{figure}

\begin{figure}[!h]
    \centering
    \includegraphics[scale=0.7]{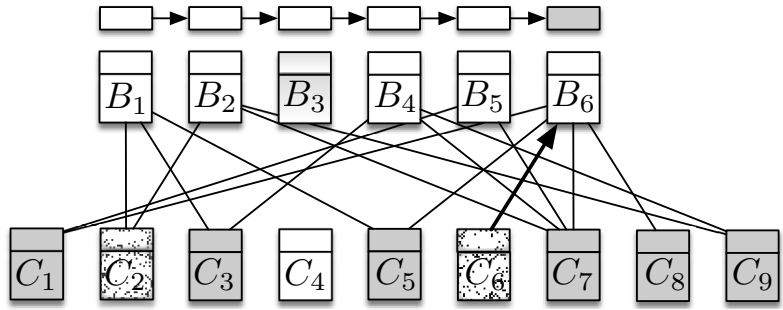}
    \caption{Iteration 2 with the bipartite graph $G^1$: the decoder rejects $\cblockj{6}$.}
    \label{fig:peeler-3}
\end{figure}

Consider the first iteration.
The decoder begins with finding a droplet, called singleton, that is connected to exactly one node in $G^{0} = G$. The only singleton in $G^{0}$ is $\cblockj{4}$, and is connected to $\block{3}$ (see Fig.~\ref{fig:peeler-2}). 
The decoder then compares the header of $\cblockj{4}$ with $\header{3}$ from the header-chain, and then verifies whether the Merkle root of the payload of $\cblockj{4}$ matches with the Merkle root stored in $\header{3}$. Since $\cblockj{4}$ is clear, the decoder will accept it (see Proposition~\ref{prop:deleting-droplet} in Appendix~\ref{app:proof-of-lemma}), and decodes $\hblock{3} = \cblockj{4}$. Then, it XORs $\cblockj{4}$ to the neighbors of $\block{3}$ excluding $\cblockj{4}$, namely $\cblockj{1}$, $\cblockj{2}$, $\cblockj{6}$, and $\cblockj{8}$. (In subsequent iterations, we refer to this step as {\it updating} the other neighbors of a decoded block.) It then removes the edges from $\block{3}$ to obtain $G^{1}$ as shown in Fig.~\ref{fig:peeler-3}.

\begin{figure}[!h]
    \centering
    \includegraphics[scale=0.7]{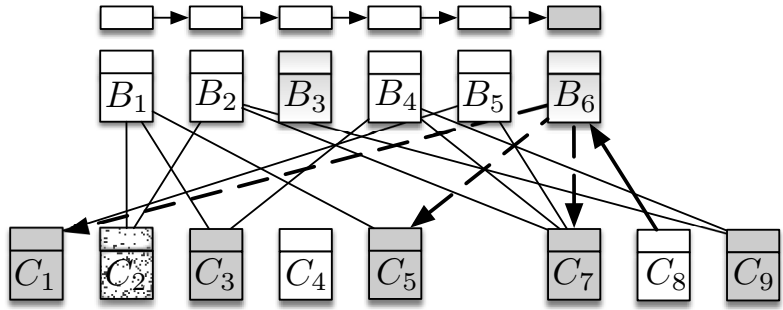}
    \caption{Iteration 3 with the bipartite graph $G^2$: the decoder accepts $\cblockj{8}$ and decodes $\block{6}$.}
    \label{fig:peeler-4}
\end{figure}

In iteration 2, there are two singletons $\cblockj{6}$ and $\cblockj{8}$. Suppose the decoder selects $\cblockj{6}$. Since the droplet is murky, the matching fails for either the header or the Merkle root (or both), and the decoder rejects $\cblockj{6}$ (see Proposition~\ref{prop:deleting-droplet}). It deletes $\cblockj{6}$ along with its edge from $G^1$ to obtain $G^2$ as shown in Fig.~\ref{fig:peeler-4}.

\begin{figure}[!h]
    \centering
    \includegraphics[scale=0.7]{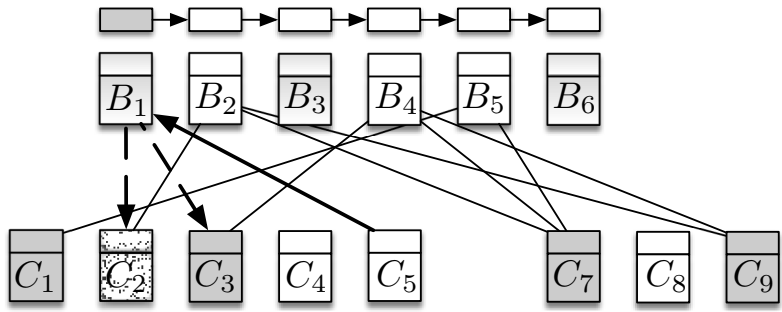}
    \caption{Iteration 4 with the bipartite graph $G^3$: the decoder accepts $\cblockj{5}$ and decodes $\block{1}$.}
    \label{fig:peeler-5}
\end{figure}

In iteration 3, the only singleton droplet is $\cblockj{8}$ that is connected to $\block{6}$. Since the droplet is clear, the headers and the Merkle roots would match. The decoder accepts $\cblockj{8}$ and decodes $\hblock{6} = \cblockj{8}$. It updates the other neighbors of $\block{6}$, and removes the edges from $\block{6}$ to obtain $G^{3}$ as shown in Fig.~\ref{fig:peeler-5}.

\begin{figure}[!h]
    \centering
    \includegraphics[scale=0.7]{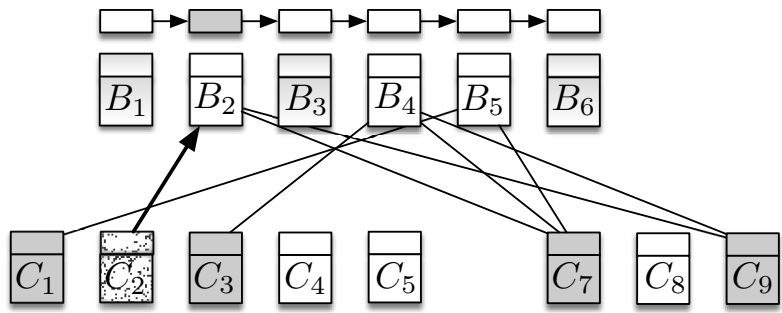}
    \caption{Iteration 5 with the bipartite graph $G^4$: the decoder rejects $\cblockj{2}$.}
    \label{fig:peeler-6}
\end{figure}

In iteration 4, there are two singletons $\cblockj{1}$ and $\cblockj{5}$. Suppose the decoder selects $\cblockj{5}$. Since the droplet is clear, the headers and the Merkle roots would match. The decoder accepts $\cblockj{5}$ and decodes $\hblock{1} = \cblockj{5}$. It updates the other neighbors of $\block{1}$, removes the edges from $\block{1}$ to obtain $G^{4}$ as shown in Fig.~\ref{fig:peeler-6}.

\begin{figure}[!h]
    \centering
    \includegraphics[scale=0.7]{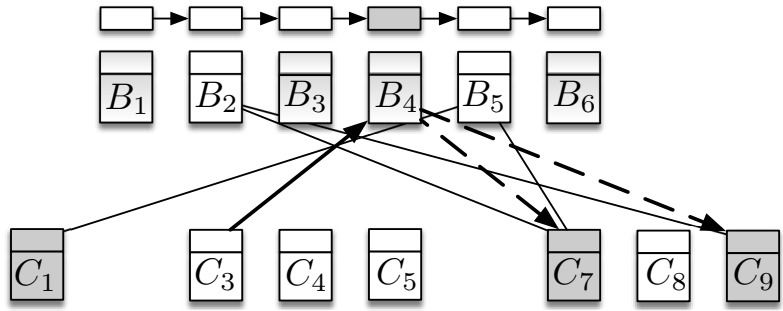}
    \caption{Iteration 6 with the bipartite graph $G^5$: the decoder accepts $\cblockj{3}$ and decodes $\block{4}$.}
    \label{fig:peeler-7}
\end{figure}

In iteration 5, there are three singletons $\cblockj{1}$, $\cblockj{2}$, and $\cblockj{3}$. Suppose the decoder selects $\cblockj{2}$. Since the droplet is murky, the matching fails for either the header or the Merkle root (or both), and the decoder rejects $\cblockj{2}$. It deletes $\cblockj{2}$ to obtain $G^5$ as shown in Fig.~\ref{fig:peeler-7}.

\begin{figure}[!h]
    \centering
    \includegraphics[scale=0.7]{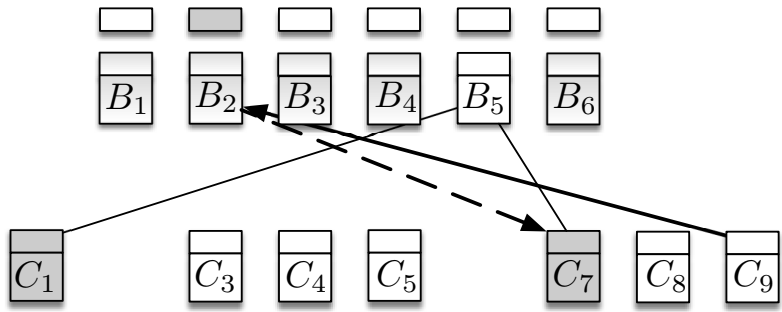}
    \caption{Iteration 7 with the bipartite graph $G^6$: the decoder accepts $\cblockj{9}$ and decodes $\block{2}$.}
    \label{fig:peeler-8}
\end{figure}

In iteration 6, out of the two singletons $\cblockj{1}$ and $\cblockj{3}$, suppose the decoder selects $\cblockj{3}$. Since the droplet is clear, the headers and the Merkle roots will match. The decoder accepts $\cblockj{3}$ and decodes $\hblock{5} = \cblockj{3}$. It updates the other neighbors of $\block{5}$, and removes the edges from $\block{5}$ to obtain $G^{6}$ as shown in Fig.~\ref{fig:peeler-8}.

\begin{figure}[!h]
    \centering
    \includegraphics[scale=0.7]{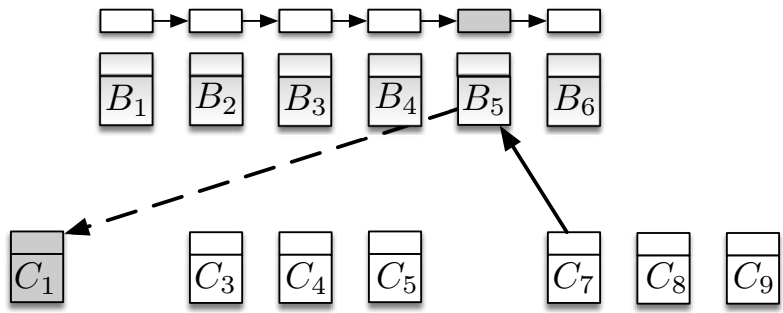}
    \caption{Iteration 8 with the bipartite graph $G^7$: the decoder accepts $\cblockj{7}$ and decodes $\block{5}$.}
    \label{fig:peeler-9}
\end{figure}

In iteration 7, the decoder chooses the singleton $\cblockj{9}$. It accepts it, and decodes $\hblock{2} = \cblockj{9}$. It updates the other neighbors of $\block{2}$. The graph $G^7$ after removing edges from $\block{2}$ is shown in Fig.~\ref{fig:peeler-9}.

Finally, iteration 8, the the decoder chooses the singleton $\cblockj{7}$. It accepts it, and decodes $\hblock{5} = \cblockj{7}$. As all the 6 blocks are decoded, the decoder stops.

\noindent {\bf Decoding Failure:} As we will show in Sec.~\ref{sec:analysis-SeF}, when a bucket node contacts a set of droplet nodes that contains slightly more that $k/s$ honest nodes, it can successfully decode the original blockchain with high probability.
However, if the decoder cannot decode from the $ns$ droplets collected by a bucket node, the node can simply contact additional droplet nodes to collect more droplets until it finds a singleton. In particular, the bucket node contacts $\hat{n}$ additional droplet nodes for some $\hat{n}\ll n$ (which can be a predetermined parameter). Arbitrarily label the downloaded droplets as $\cblockj{ns+1}$, $\cblockj{ns+2}$, $\ldots$, $\cblockj{(n+\hat{n})s}$. First, remove the contribution of already decoded blocks from each of $\cblockj{j}$. Specifically, if block a block, say $\block{m}$, used in computing $\cblockj{j}$ is already decoded, then update $\cblockj{j}$ as $\cblockj{j} \leftarrow \cblockj{j} \oplus \hblock{m}$. Next, append these updated droplets as right vertices in $G^{i-1}$. Add an edge connecting a droplet $\cblockj{j}$, $ns+1 \leq j \leq (n+\hat{n})s$, to block $\block{m}$ if $\block{m}$ is not yet decoded and it is used in computing $\cblockj{j}$. If there is a singleton amongst the newly downloaded droplets, then proceed to Step (3). Otherwise, contact $\hat{n}$ additional droplet nodes. The decoder declares failure when the bucket node is unable to find additional droplet nodes.

\noindent {\bf Turning from a bucket node to a droplet node:} After the bucket node decodes the original blockchain, it computes its own droplets for every epoch by following the encoder in Sec.~\ref{sec:encoding}. At this point, the bucket node turns itself into a droplet node which, in turn, can help a new bucket node. 


\subsubsection{Degree Distribution}
\label{sec:degree-distribution}

While the encoder and the decoder are valid for any degree distribution, the probability of successfully decoding the input symbols (the blockchain in our case) from a given number of output symbols (droplets in our case) depends on the choice of the degree distribution. In the following, we describe the {\it robust soliton degree distribution} proposed by Luby~\cite{Luby:02}. The robust soliton degree distribution is shown to have good probability of success (without any adversarial nodes) in~\cite{Luby:02}.

Recall that a {\it degree distribution} $\mu(\cdot)$ is a discrete probability mass function on integers between $1$ and $k$. 
In order to describe the robust soliton degree distribution, we introduce the following notation. 
First, define a function $\rho(\cdot)$ as:\footnote{It is not hard to verify that $\sum_{d=1}^{k}\rho(d) = 1$, and thus, $\rho$ is a probability distribution. This distribution is referred to as the {\it ideal soliton distribution}. For further details, and to understand why it is called ``soliton'', we refer the reader to~\cite[Sec.~3.2]{Luby:02}.}
\begin{equation}
    \rho(d) = 
    \begin{cases}
    \frac{1}{k} & \textrm{for}\:\: d = 1\\*
    \frac{1}{d(d-1)} & \textrm{for}\:\: d = 2,\ldots,k.
    \end{cases}
\end{equation}
Next, for given $0 < \delta <1$ and $c>0$, define 
\begin{equation}
    \label{eq:R-for-LT-codes}
    R = c\sqrt{k}\ln\left(\frac{k}{\delta}\right).
\end{equation} 
Further, define a function $\theta(\cdot)$ as
\begin{equation}
    \theta(d) = 
    \begin{cases}
    \frac{R}{dk} & \textrm{for}\:\: d = 1,\ldots,k/R-1\\
    \frac{R}{k}\ln\left(\frac{R}{\delta}\right) & \textrm{for}\:\: d = k/R\\
    0 & \textrm{for}\:\: d = k/R+1,\ldots,k.
    \end{cases}
\end{equation}
As we will see in Sec.~\ref{sec:analysis}, the parameter $\delta$ gives a (conservative) bound on the probability that the decoding fails to succeed after a certain number of droplets are downloaded. The parameter $c$ is a free parameter that can be tuned to optimize the number of droplets required to recover the blockchain. Adding  $\rho(\cdot)$ to $\theta(\cdot)$ and normalizing gives the {\it robust soliton distribution} as:
\begin{equation}
    \label{eq:robust-soliton}
        \mu(d) = \frac{\rho(d) + \theta(d)}{\beta}, \quad \textrm{for}\:\: d = 1,\ldots,k,
\end{equation}
where 
\begin{equation}
    \label{eq:beta}
    \beta = \sum_{d=1}^{k}\rho(d)+\theta(d).
\end{equation}

\section{Performance Analysis}
\label{sec:analysis}

We begin with formally defining the performance metrics that were briefly described in Sec.~\ref{sec:model}. Consider a coding scheme with a pair of encoding and decoding schemes $(\texttt{Enc},\texttt{Dec})$. 

\noindent
{\bf Storage Savings} of a coding scheme is the ratio of the total blockchain size to the size of the encoded blockchain. 
Specifically, the storage savings of a droplet node $j$ is $\frac{\size{B}}{\size{\Enc{B,j}}}$. 

\noindent
{\bf Bootstrap Cost:} Consider a coding scheme with a storage savings of $\gamma$. For a given $0 < \delta < 1$, the bootstrap cost of a coding scheme is measured by the minimum number of honest droplet nodes $K(\gamma, \delta)$ that a bucket node needs to contact in order to ensure that the blockchain can be recovered with probability at least $1-\delta$. 

Note that $K(\gamma, \delta)$ can be considered as the minimum number of honest droplet nodes that the system must contain to guarantee, with probability at least $1-\delta$, that the historical blockchain data is preserved. Thus, the bootstrap cost of a coding scheme reflects the {\it security performance} of the system. 
    
\noindent
{\bf Bandwidth Overhead} is the overhead in terms of the amount of data that a bucket node needs to download for recovering the blockchain with high probability. Specifically, the bandwidth overhead is computed as the amount of data required to be downloaded for ensuring successful blockchain recovery minus the size of the blockchain (at the time of bootstrap) divided by the size of the blockchain. For a coding scheme with storage savings of $\gamma$, wherein a $\sigma$-fraction of droplet nodes are malicious and the blockchain should be recovered with probability at least $1 - \delta$, we denote the bandwidth overhead by $\bandwidth{\gamma,\delta,\sigma}$. 

     
\noindent
{\bf Computation Cost} of a coding scheme is measured in terms of the average number of arithmetic operations associated with the encoder $\texttt{Enc}$ and the decoder $\texttt{Dec}$.
In particular, the encoding cost is the expected number of arithmetic operations sufficient for generating droplets, divided by the number of original blocks. Similarly, the decoding cost is the expected number of arithmetic operations sufficient to recover the blockchain, divided by the number of original blocks. 

\noindent {\bf Decentralization:} a droplet node should be able to generate its droplets without knowing what any other node in the system is storing.

We begin with establishing a fundamental trade-off between the storage savings and bootstrap cost for any coding scheme. 
For simplicity, we focus our attention to coding schemes in which each droplet node achieves the storage savings of $\gamma$.\footnote{When this is not the case, using the similar proof as that of~\ref{prop:lower-bound}, it is easy to show that the lower bound on the security performance for a coding scheme is $\gamma_{min}$, which is the minimum storage savings achieved by the scheme.}

\begin{theorem}
\label{prop:lower-bound}
For any $0\leq \delta < 1$, the bootstrap cost of any coding scheme in which each droplet node achieves the storage savings of $\gamma$ is lower bounded by $\lceil\gamma\rceil$, \ie, $K(\gamma,\delta) \geq \lceil\gamma\rceil$.
\end{theorem}
\begin{proof}
Suppose that there exist $n$ honest droplet nodes from which it is possible to recover the blockchain. In order to recover the blockchain, the total size of the downloaded data must be at least the size of the blockchain. Further, each of the $n$ droplet nodes can contribute $\size{B}/\gamma$ amount, since every droplet node is achieving the storage savings of $\gamma$. Therefore, $n$ should be at least $\lceil\gamma\rceil$. 
\end{proof}
\noindent Note that the above theorem implies that the network must contain at least $\lceil\gamma\rceil$ honest droplet nodes to guarantee that the historical blockchain data is preserved.

\subsection{SeF Codes}
\label{sec:analysis-SeF}

First, we show that SeF codes guarantee that the blockchain can be  successfully recovered with overwhelming probability as long as the set of droplet nodes contacted by a bucket node contains sufficiently many honest droplet nodes. 
Towards this end, we assume that droplet nodes randomly sample degrees and neighbors for computing the droplets in the first epoch (see Step (2) of the encoder), and then use the same degree and neighbors in subsequent epochs.\footnote{As we will see in the proof of Lemma~\ref{lem:decoding-condition}, this assumption ensures that if a bucket node can (resp. cannot) recover the blocks in the first epoch, it can (resp. cannot) recover all (resp. any of) the subsequent epochs.}

\begin{lemma}
\label{lem:decoding-condition}
Consider a bucket node that contacts an arbitrary set of droplets nodes during its bootstrap. If this set contains at least $\frac{1}{s}\left(k + O\left(\sqrt{k}\ln^2(k/\delta)\right)\right)$ honest droplet nodes, then the probability that the error-resilient peeling decoder fails to recover the entire blockchain is at most $\delta$.
\end{lemma}
\begin{proof}
The proof is deferred to Appendix~\ref{app:proof-of-lemma}. 
\end{proof}
\noindent The above lemma implies that successful blockchain recovery is guaranteed with high probability as long as the network contains $\frac{1}{s}\left(k + O\left(\sqrt{k}\ln^2(k/\delta)\right)\right)$ honest droplet nodes. In other words, SeF codes can ensure that the blockchain history is preserved even if an adversary corrupts a large fraction of droplet nodes.

Next, we analyze the performance of SeF codes. 

\noindent{\bf Assumptions:} We make the following assumptions to simplify the analysis. 
\begin{itemize}
	\item[(i)] While characterizing the storage savings, we assume that the storage space required to store the binary vector $\vj{j}$ corresponding to a droplet is negligible as compared to the size of the droplet. Note that storing a length-$k$ binary vector requires only $log_2(k)$ bits; \eg, for $k = 10000$, it takes only 1250 bytes. Thus, for large enough block size (\eg, 1MB), this assumption can be justified. Further, we assume that the storage space required to store the header-chain and the blocks in the current epoch is negligible as compared to the size of the blockchain. Note that, since the blockchain is an ever-growing data structure, this assumption can be easily justified.
	\item[(ii)] While characterizing the bandwidth overhead, we assume that, if a $\sigma$-fraction of droplet nodes are malicious, then a droplet node contacted by a bucket node turns out to be malicious with probability $\sigma$, independent of the other contacted nodes. Here, we implicitly assume that a bucket node can contact a random subset of droplet nodes. This is because, in any protocol, malicious nodes can induce heavy bandwidth overhead by {\it surrounding} a bucket node, say by hijacking its connections, and by providing {\it garbage} data. Therefore, assuming that a bucket node can contact a random subset of droplet nodes allows us to obtain average bandwidth overhead.
	\item[(iii)] While characterizing the computation cost associated with decoding, we do not include the number of arithmetic operations required to compute a Merkle root in Step 3(a). This is because a node anyway needs to compute the Merkle root in order to validate a block. 
\end{itemize}

\begin{theorem}
\label{thm:SeF-performance}
SeF codes are decentralized and achieve the following performance measures:
\begin{enumerate}
    \item Storage savings: $\gamma = k/s$;
    \item Bootstrap cost: $\K(k/s,\delta) = \frac{k + O\left(\sqrt{k}\ln^2(k/\delta)\right)}{s}$;
    \item Bandwidth overhead: $\bandwidth{k/s,2\delta,\sigma} = O\left(\frac{\ln^2(k/\delta)}{(1-\sigma)\sqrt{k}}\right)$;
    \item Computation cost: encoding cost = $O\left(\frac{s\ln(k/\delta)}{k}\right)$, decoding cost = $O\left(\frac{\ln(k/\delta)}{1-\sigma}\right)$.
\end{enumerate}
\end{theorem}
\begin{proof}
The proof is deferred to Appendix~\ref{app:proof-of-theorem-1}.
\end{proof}

\noindent We can immediately make the following observations about the performance of SeF codes.
\begin{remark}
\label{rem:observations}
First, observe that the bootstrap cost for SeF codes is off from its optimal value of $\frac{k}{s}$ (see Theorem~\ref{prop:lower-bound}) by $\frac{O\left(\sqrt{k}\ln^2(k/\delta)\right)}{s}$. In other words, the overhead with respect to the optimal bootstrap cost is $\frac{O\left(\ln^2(k/\delta)\right)}{\sqrt{k}}$, which goes to zero as $k$ increases. Next, observe that the bandwidth overhead also goes to zero as $k$ increases. In fact, it is easy to see that the bandwidth overhead (resp. bandwidth cost) is proportional to the bootstrap overhead (resp. bootstrap cost). This essentially follows from all the blocks, and hence, all the droplets having the same size. On the other hand, in a practical blockchain, bandwidth overhead is no longer proportional to bootstrap overhead due to variability in block size as we will see in our experiments (Sec.~\ref{sec:simulations}). Finally, the normalized encoding cost goes to zero with $k$, while the normalized decoding costs grows logarithmically in $k$.
\end{remark}

\subsection{Random Sampling and Reed-Solomon Codes}
\label{sec:analysis-others}

{\bf Random Sampling:} In this simple scheme, in each epoch of length $k$, a droplet node stores $s$ distinct blocks that are selected uniformly at random.\footnote{It is worth noting that a similar scheme is used in the Ripple blockchain, and is referred to as {\it history sharding}~\cite{Ripple-sharding:18}. In history sharding, the transaction history of the XRP Ledger is partitioned into segments, called shards. A server that has enabled history sharding acquires and stores randomly selected shards.} Note that this scheme achieves the storage savings of $k/s$, since the storage grows by $s$ blocks when the blockchain grows by $k$ blocks. 

As noted in~\cite{Luby:02}, random sampling can be considered as a special case of LT codes for the following degree distribution (referred to as {\it all-at-once} distribution).
\begin{equation}
\label{eq:all-at-once}
    \rho(d) = 
    \begin{cases}
    1 & \textrm{if}\:\: d = s\\*
    0 & \textrm{otherwise}.
    \end{cases}
\end{equation}

Even though random sampling has trivial encoding and decoding costs, its
major limitation is that it incurs a significant bootstrap cost. To see this, consider $s = 1$ for simplicity, and focus on the first epoch. It is easy to see that recovering the blockchain in this case is equivalent to the classical ``coupon collector'' problem (see, \eg,~\cite[Chapter 3.6]{Motwani:95}), which incurs a (multiplicative) logarithmic hit in bootstrap cost.\footnote{It is worth noting that, for $s > 1$, the random sampling scheme is equivalent to the coupon collector with group drawing problem, and the analysis is similar, see,~\eg,~\cite{Stajde:90}.} In particular, it is necessary to contact $k\ln(k/\delta)$ honest droplet nodes on average in order to recover the blockchain with probability at least $1-\delta$. 

\vspace{2mm}
\noindent {\bf Reed-Solomon (RS) Codes:} We begin with the following notation. Let $\GF{q}$ denote the Galois field of size $q$. Note that, when the maximum size of a block is $L$ bits, every block can be considered as an element of $\GF{2^L}$. Consider $L' \geq L$ such that $L$ divides $L'$. Then, $\GF{2^{L'}}$ is an extension field of $\GF{2^L}$. For simplicity, we assume that $L' = \Omega(\log_2(N))$, where $N$ denotes the total number of droplet nodes in the network.

Now, we describe the encoding procedure for an RS code, focusing on the first epoch. A droplet node samples $s$ points from $\GF{2^{L'}}$ uniformly at random, and stores the evaluations the following polynomial $B(x)$ on these points: $B(x) = \block{1} + \block{2}x + \cdots + \block{i}x^{i-1} + \block{k}x^{k-1}$, where $\block{1},\ldots,\block{k}$ are the first $k$ blocks.
Note that it is possible to interpolate $B(x)$ from its evaluations on any $k$ distinct points. Further, for a large enough $2^{L'}$, an arbitrary set of $k/s$ honest droplet nodes will have evaluations of $B(x)$ on $k$ distinct points with high probability. Therefore, an RS code allows a bucket node to decode the blockchain (with high probability) from any $k/s$ honest droplet nodes via polynomial interpolation. Hence, an RS code achieves the {\it optimal} bootstrap cost of $k/s$. 
Moreover, as long as the network contains $k/s$ honest droplet nodes, it is possible, in principle, to recover the blockchain. 

However, recovering the blockchain when the network contains a small number of honest nodes will require heavy computation cost. To see this, consider the case when the network contains exactly $k/s$ honest droplet nodes. Since a bucket node cannot distinguish an honest droplet node from a malicious one just by observing its stored droplets, it needs to employ the following decoding strategy. First, it contacts an arbitrary subset of $k/s$ droplet nodes, and downloads their droplets. Using these droplets, it  recovers a candidate blockchain via polynomial interpolation, and checks the validity of the recovered blockchain using the header chain. If the validity fails, it contacts another subset of $k/s$ droplet nodes and repeats the procedure. In the worst case, the node may need to contact every $(k/s)$-subset of droplet nodes, resulting in a prohibitive computation cost. 

In practice, one can use algorithms designed to decode RS codes in the presence of errors, \eg, Peterson-Gorenstein-Zierler algorithm~\cite{MacWilliams-Slone:78}. The best known computation cost for decoding a length-$N$ RS code is $O(N polylog(N))$, see, \eg~\cite{Didier:RS:09}. 
Note that algorithms designed to decode RS codes in the presence of errors do not need to use the header chain as a side-information. However, such algorithms can tolerate only $\frac{(Ns - k)}{2}$ adversarial droplets among $Ns$ droplets. Thus, the blockchain can be recovered only when the network contains at least $\left(\frac{N}{2}+\frac{k}{2s}\right)$ honest droplet nodes, requiring more than half of the droplet nodes to be honest. 

\section{Practical Issues}
\label{sec:practical}

\subsection{Tackling Variability in Block Size}
\label{sec:block-sizes}
Until now, we have assumed that all the blocks have the same size. On the other hand, popular blockchains such as Bitcoin and Ethereum produce blocks with variable size (see~\cite{Bitcoin-blocksize} and~\cite{Ethereum-blocksize}, respectively). In this section, we discuss how to handle variability in block size. 

In a blockchain with a limit on the block size, the simplest way to deal with variable block sizes is to zero pad every block to the maximum size during encoding. However, when the average block size is smaller than the maximum, this results in higher storage costs. In the following, we discuss two simple and efficient protocols to handle variable block size.

\begin{itemize}
    \item[1.]  {\bf Adaptive zero-padding:} Recall that in LT encoding a node first chooses a degree $d$ using a degree distribution. Then, it chooses $d$ distinct blocks from the epoch under consideration. Then, while computing the bit-wise XOR, the node can simply zero-pad the blocks to the largest block among the $d$ blocks. We refer to this procedure as adaptive zero-padding. 
    
    Adaptive zero-padding performs well when the variance in  block size is small. However, it can perform poorly when the variance in block size is large. To overcome this issue, we propose to concatenate several contiguous blocks in the following. 
    
    \item[2.] {\bf Block Concatenation:} A natural way to reduce variance in block size is to first concatenate blocks to form {\it super-blocks} of approximately same size, and then perform encoding on the super-blocks.  
        More specifically, let $L$ denote the maximum block size, and let $\Ls \geq L$ be a design parameter. For example, for the Bitcoin blockchain with $L = 1MB$, we use $\Ls = 1, 5,\:\: \textrm{and}\:\: 10MB$ in our simulations. For two binary strings $\block{i}$ and $\block{j}$, let $\block{i}\mid\mid\block{j}$ denote their concatenation.
        For simplicity, we assume that the block header contains the size of the block.
    
    {\it Block concatenation procedure:}
    \begin{itemize}
        \item[(i)] {\it Initialization:} Set super-block count $j = 1$ and block count $i = 1$.
        \item[(ii)] Set super-block $\bar{B}_j = \texttt{NULL}$. 
        
        \begin{itemize}
            \item[a.] If $\size{\bar{B}_j \mid\mid B_i} \leq \Ls$,  
        
        $\quad$ Set $\bar{B}_j \leftarrow \bar{B}_j \mid\mid B_i$. 
        
        $\quad$ Increment $i$.
        
        $\quad$ Go to Step (ii)-a.
        
        \item[b.] Else,
        
        $\quad$ Increment $j$.
        
        $\quad$ Go to Step (ii).
        \end{itemize}
    \end{itemize}
    
    We define an epoch as the time required for the blockchain to grow by $k$ super-blocks. The actual number of blocks produced in an epoch will vary depending on the block sizes. LT encoding is performed on super-blocks. For instance, in the first epoch, LT encoding is then performed on super-blocks $\bar{B}_1$, $\bar{B}_2$, $\cdots$, $\bar{B}_k$. Note that the encoder may still need to use adaptive zero padding while XORing super-blocks. However, the size of a super-block is at least $\Ls - L$. Thus, choosing $\Ls$ to be sufficiently larger than $L$ ensures small variance in super-block sizes, reducing the overhead incurred by adaptive zero padding. 
        
    In the error-resilient peeling decoder in Sec.~\ref{sec:decoding}, we modify Step 3 to check all the blocks in a singleton super-block. To be more precise, consider Step 2 in  at which the bucket node finds a singleton super-block, say $\cblockj{l}$. Assuming that the header contains the block size, the bucket node knows from the header chain that the $l$-th super-block should be a concatenation of blocks $i+1$ to $i+p$ for some $i \geq 0$ and $p \geq 0$. In other words, if $\cblockj{l}$ is a clear droplet, then it will have the following structure: $\cblockj{l} = \{\{H_{i+1},T_{i+1}\}, \{H_{i+2},T_{i+2}\}, \cdots, \{H_{i+p},T_{i+p}\}\}$ for some $i$ and $p$. 
    
    Assuming that the headers have the same size and the block-size is included in the header, it is possible to decompose $\cblockj{l}$ in the following form: $\cblockj{l} = \{\{\hat{H}_{l_1},\hat{T}_{l_1}\}, \{\hat{H}_{l_2},\hat{T}_{l_2}\}, \cdots, \{\hat{H}_{l_p},\hat{T}_{l_p}\}\}$. Then, in the Step 3, the singleton $\cblockj{l}$ is accepted only if, for each $1\leq j\leq p$, $\hat{H}_{l_j}$ matches with $H_{i+j}$ and  $\mroot{\hat{T}_{l_j}}$ matches with the Merkle root in $H_{i+j}$. Otherwise, the singleton is rejected. The rest of the decoding algorithm remains the same.
    
\end{itemize}

\subsection{Obtaining the Honest Header-Chain}
\label{sec:header-chain}
While describing the error-resilient peeling decoder, we assumed that a bucket node has an access to the honest (correct) header-chain. It is easy for a bucket node to obtain the correct header chain. In particular, a bucket node can simply query a large number of droplet nodes to obtain the longest {\it valid}\footnote{A header-chain is said to be valid if it follows the hash-chain structure, and proof-of-work puzzles are correctly solved.} header-chain. Note that even though the error-resilient peeling decoding is performed separately for each epoch, a node obtains a copy of the longest valid header-chain up to the current height. Assuming that the majority of the mining power is honest and the adversary has limited computing power, the longest valid header-chain is the correct header chain with overwhelming probability. Thus, as long as the bucket node can contact one honest droplet node, it is guaranteed to obtain the correct header-chain. 

It is worth noting that light (also called SPV or thin) clients, which are an integral part of several practical blockchain protocols like Bitcoin and Ethereum, are designed to obtain the longest header-chain; see, \eg~\cite{Bitcoin-wiki-spv,Ethereum-wiki-light-client}. Thus, a bucket node can first act as a light client before starting to collect the droplets.

\section{Simulation Results}
\label{sec:simulations}

We begin with numerical analysis of the performance of the proposed SeF codes. 
Without loss of generality, we consider the first epoch.  We consider the following set of parameters for LT codes (cf.~\eqref{eq:R-for-LT-codes}): $c = \{0.01, 0.03, 0.1, 0.3\}$ and $\delta = \{0.1, 0.3, 0.5, 0.7\}$. We choose the values of $c$ and $\delta$ that result in the best performance. For any setup that we consider, the experiments are repeated 100 times to compute the statistics. 

\begin{figure}[!t]
    \centering
    \begin{subfigure}[t]{0.5\textwidth}
        \centering
        \includegraphics[width=\textwidth]{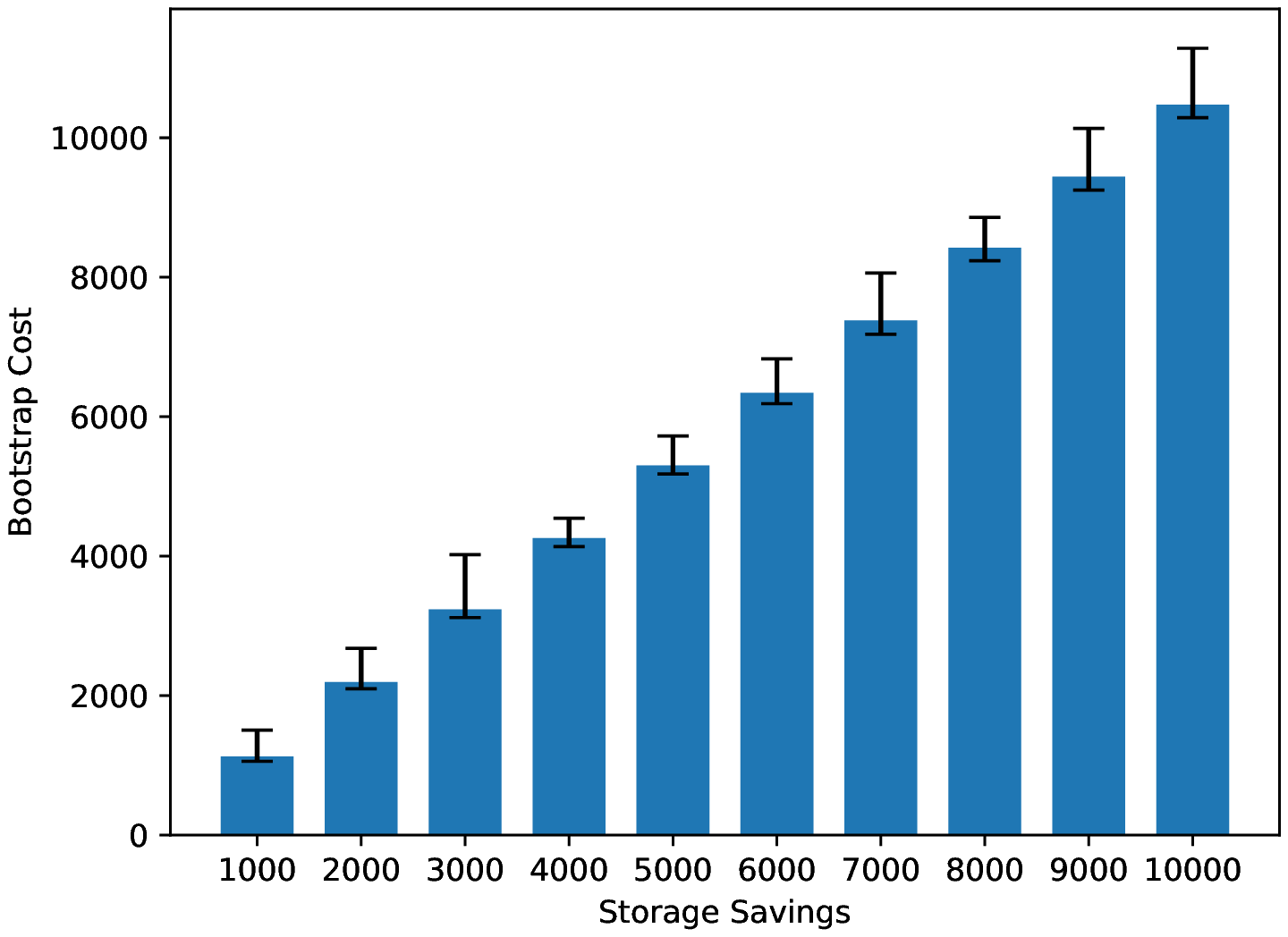}
        \caption{SeF Codes}
        \label{fig:bootstrap-cost-bar-LT}
    \end{subfigure}%
    ~ 
    \begin{subfigure}[t]{0.5\textwidth}
        \centering
        \includegraphics[width=\textwidth]{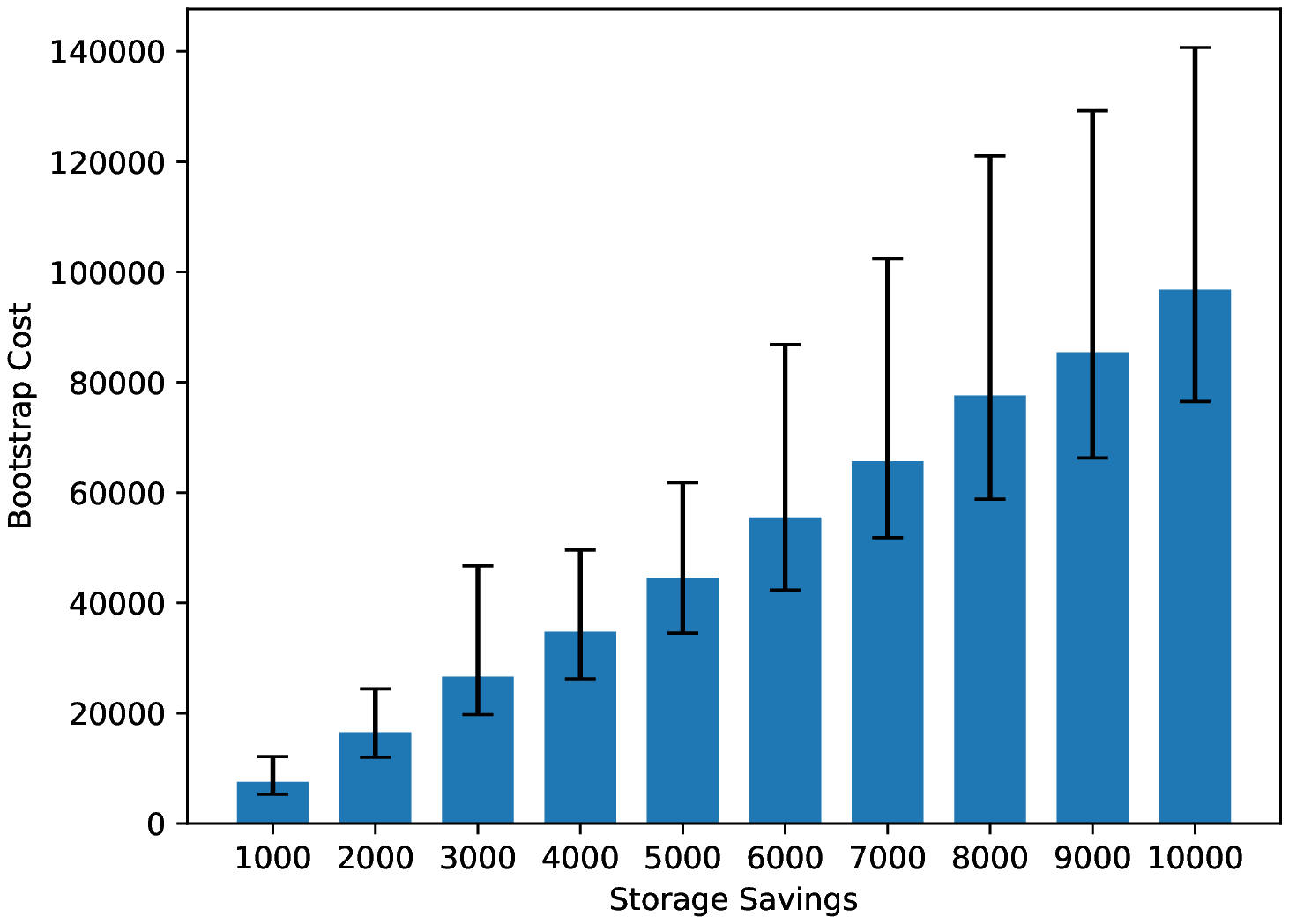}
        \caption{Random Sampling}
        \label{fig:bootstrap-cost-bar-CC}
    \end{subfigure}
    \caption{{\it Average bootstrap cost versus storage savings.}}
    \label{fig:bootstrap-cost-bar}
\end{figure}

\begin{figure}[!t]
    \centering
    \includegraphics[scale=0.65]{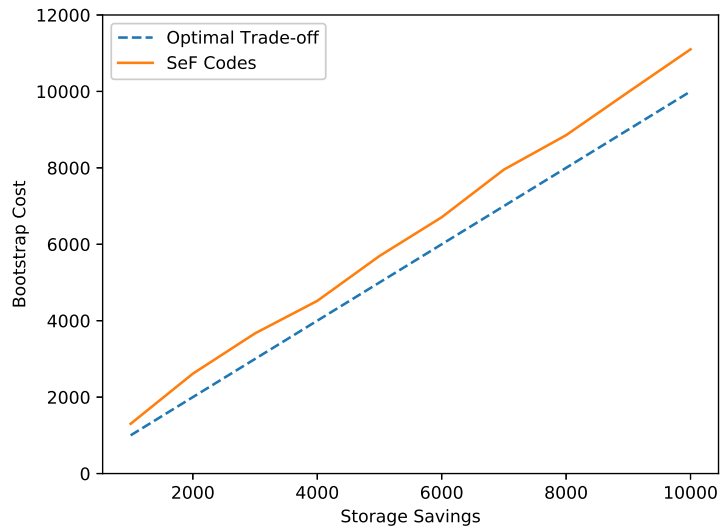}
    \caption{{\it Bootstrap cost versus storage savings to ensure successful blockchain recovery with $99\%$.}}
    \label{fig:cost-vs-savings-2}
\end{figure}

First, we plot the average bootstrap cost versus storage savings for SeF codes in Fig.~\ref{fig:bootstrap-cost-bar-LT}. We also plot the minimum and maximum bootstrap cost over 100 trials. Observe that, for a given storage savings of $\gamma$, the bootstrap cost of SeF codes is close to the optimum bootstrap cost $\gamma$. For comparison, we plot the bootstrap cost versus storage savings for random sampling in Fig.~\ref{fig:bootstrap-cost-bar-CC}.
To highlight that SeF codes achieve near optimum trade-off between the bootstrap cost and the storage savings, we plot the bootstrap cost that ensures successful blockchain recovery with $99\%$ in Fig.~\ref{fig:cost-vs-savings-2} along with the optimal bootstrap cost. 

%

\begin{figure}[!t]
    \centering
    \begin{subfigure}[t]{0.5\textwidth}
        \centering
        \includegraphics[width=\textwidth]{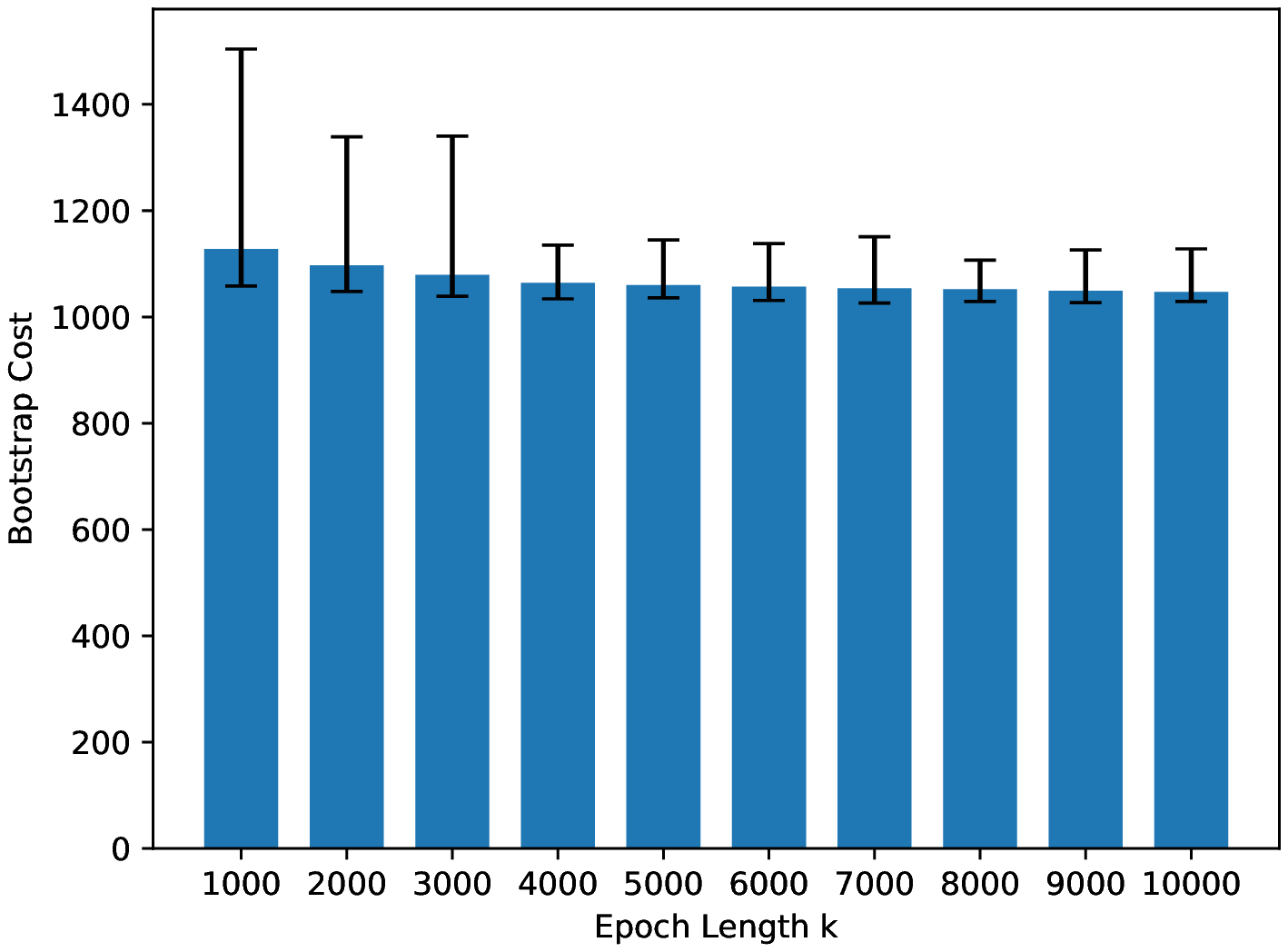}
        \caption{SeF Codes}
        \label{fig:bootstrap-cost-bar2-LT}
    \end{subfigure}%
    ~ 
    \begin{subfigure}[t]{0.5\textwidth}
        \centering
        \includegraphics[width=\textwidth]{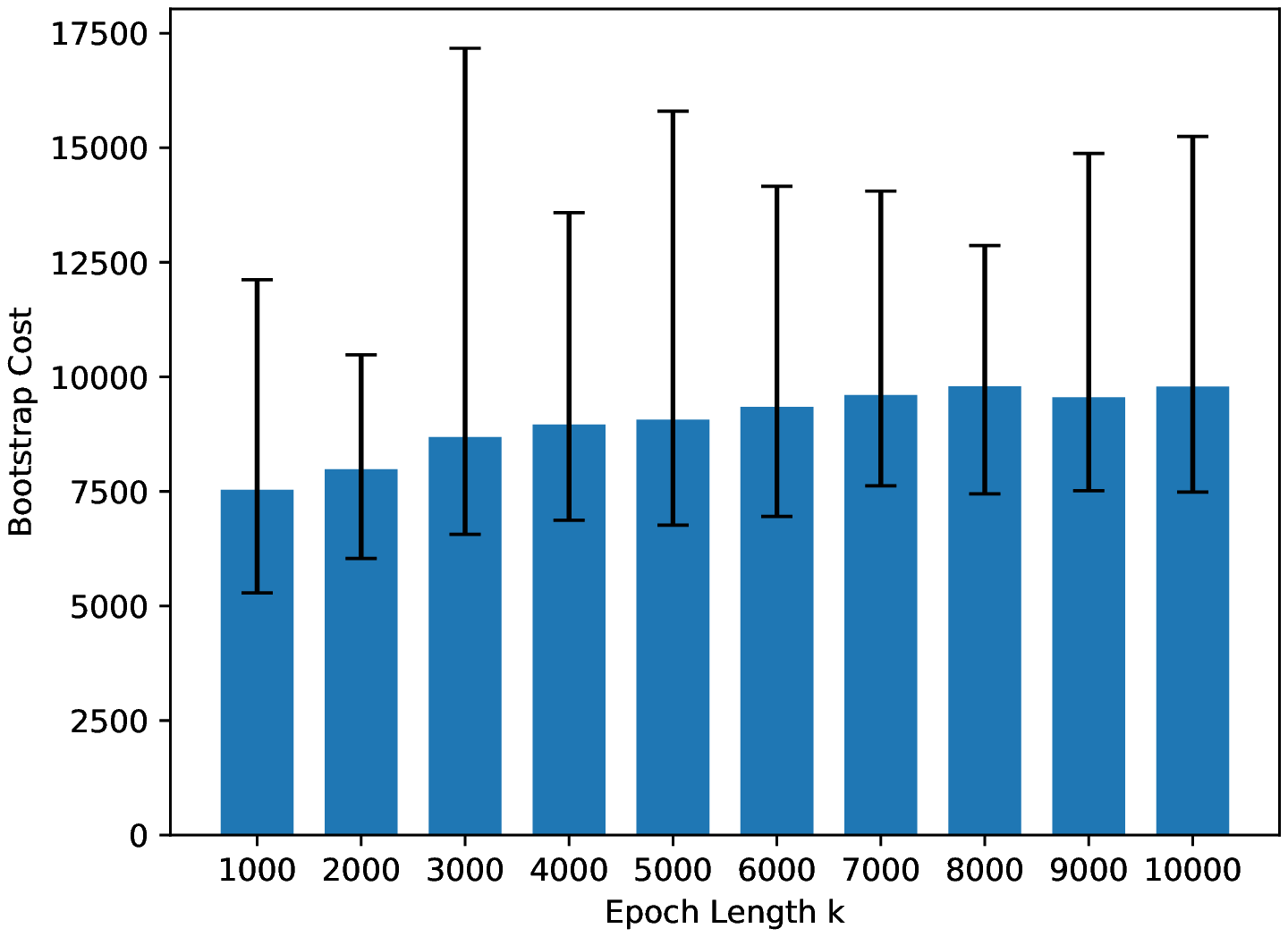}
        \caption{Random Sampling}
        \label{fig:bootstrap-cost-bar2-CC}
    \end{subfigure}
    \caption{{\it Average bootstrap cost versus epoch-length $k$.}}
    \label{fig:epoch-length}
\end{figure}

Next, we study the effect of epoch-length $k$ on the bootstrap cost in Fig.~\ref{fig:epoch-length}. In particular, we increase $k$ and $s$ such that the storage savings is $k/s = 1000$, and plot the average bootstrap cost. One can see that as the epoch length increases, the bootstrap cost for SeF codes gets closer to the optimal value of $1000$. This is because LT codes are more efficient for larger $k$. On the other hand, for a larger epoch-length $k$, a droplet node needs larger buffer space to store the blocks of the current epoch before they can be encoded. We also plot the bootstrap cost versus the epoch length for random sampling for comparison in Fig.~\ref{fig:bootstrap-cost-bar2-CC}.


\begin{figure}[!t]
    \centering
    \includegraphics[scale=0.65]{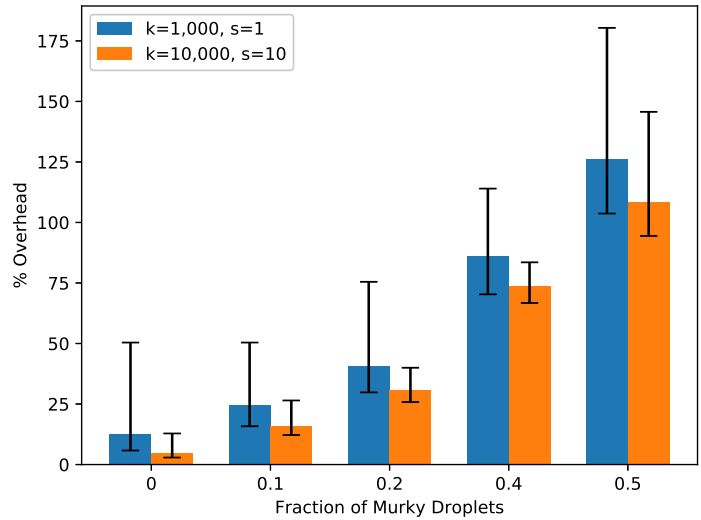}
    \caption{{\it Average bandwidth overhead versus fraction of adversarial droplet nodes for SeF codes.}}
    \label{fig:bandwidth-overhead}
\end{figure}

Next, we  plot bandwidth overhead as a function of a fraction $\sigma$ of adversarial droplet nodes in Fig.~\ref{fig:bandwidth-overhead}. Recall that we make the following assumption about the network model during the bootstrap process: if a $\sigma$-fraction of droplet nodes are malicious, then a droplet node contacted by a bucket node turns out to be malicious with probability $\sigma$. We consider two parameter settings, targeted at  $1000\times$ storage savings: (i) ($k = 1000$, $s = 1$); and (ii) ($k = 10000$, $s = 10$). Observer that $k = 10000$, $s = 10$ results in a smaller bootstrap overhead as compared to $k = 1000$, $s = 1$.

\subsection*{Simulations on the Bitcoin Blockchain}
\label{sec:bitcoin}

In this section, we describe experiments carried out on the Bitcoin blockchain. We consider two parameter settings, targeted at  $1000\times$ storage savings: (i) ($k = 1000$, $s = 1$); and (ii) ($k = 10000$, $s = 10$). 
We observe that the actual storage savings (as well as the bandwidth overhead) is affected by variability in block size. 
To tackle block size variability, we use adaptive zero padding and block concatenation as discussed in Sec.~\ref{sec:block-sizes}.  We list the average values for storage savings, bootstrap cost, and bandwidth overhead  in Tables~\ref{tbl:bitcoin-k-1000-s-1} and~\ref{tbl:bitcoin-k-10000-s-10}. (We include the details of the experimental results in Appendix~\ref{app:experiment-details}.) 

We observe that simply using adaptive zero padding does not yield a good performance, since the block size variability in the Bitcoin is significantly large. On the other hand, block concatenation successfully mitigates the block size variability. As we increase the super-block size from $1MB$ to $10MB$, the variance in the super-block size reduces, resulting in the performance improvement. 


\begin{table}[t!]
\centering
\begin{tabular}{|c|c|c|c|c|}
\hline
$k=1000$, $s=1$ & 
\begin{tabular}{@{}c@{}} 
Adaptive Zero  \\ Padding
\end{tabular} 
& 
\begin{tabular}{@{}c@{}} 
Block Concatenation \\ to 1MB
\end{tabular} 
& 
\begin{tabular}{@{}c@{}} 
Block Concatenation \\ to 5MB
\end{tabular} 
&
\begin{tabular}{@{}c@{}} 
Block Concatenation \\ to 10MB
\end{tabular} 
\\
\hline
\hline
\begin{tabular}{@{}c@{}} 
Average Storage  \\ Savings 
\end{tabular} 
& 
749.44
& 
896.06
& 
961.33
&
978.93
\\
\hline
\begin{tabular}{@{}c@{}} 
Average Bootstrap  \\ Cost
\end{tabular} 
& 
1128
& 
1128
& 
1128
&
1128
\\
\hline
\begin{tabular}{@{}c@{}} Average Bandwidth \\ 
Overhead \\
(All Honest)\end{tabular}
& 
50.58\%
& 
25.97\%
& 
17.35\%
&
15.32\%
\\
\hline
\begin{tabular}{@{}c@{}} Bandwidth \\ 
Overhead\\
(10\% Malicious)\end{tabular}
& 
67.30\%
& 
39.95\%
& 
30.35\%
&
27.97\%
\\
\hline
\end{tabular} 
\caption{Results on the Bitcoin blockchain for $k = 1000$ and $s = 1$.}
\label{tbl:bitcoin-k-1000-s-1}
\end{table}

\begin{table}[t!]
\centering
\begin{tabular}{|c|c|c|c|c|}
\hline
$k=10000$, $s=10$ & 
\begin{tabular}{@{}c@{}} 
Adaptive Zero  \\ Padding
\end{tabular} 
& 
\begin{tabular}{@{}c@{}} 
Block Concatenation \\ to 1MB
\end{tabular} 
& 
\begin{tabular}{@{}c@{}} 
Block Concatenation \\ to 5MB
\end{tabular} 
&
\begin{tabular}{@{}c@{}} 
Block Concatenation \\ to 10MB
\end{tabular} 
\\
\hline
\hline
\begin{tabular}{@{}c@{}} 
Average Storage  \\ Savings 
\end{tabular} 
& 
744.80
& 
894.47
& 
958.60
&
976.61
\\
\hline
\begin{tabular}{@{}c@{}} 
Average Bootstrap  \\ Cost
\end{tabular} 
& 
1048
& 
1048
& 
1048
&
1048
\\
\hline
\begin{tabular}{@{}c@{}} Average Bandwidth \\ 
Overhead \\
(All Honest)\end{tabular}
& 
40.69\%
& 
17.10\%
& 
9.26\%
&
7.33\%
\\
\hline
\begin{tabular}{@{}c@{}} Average Bandwidth \\ 
Overhead\\
(10\% Malicious)\end{tabular}
& 
56.38\%
& 
30.19\%
& 
21.59\%
&
19.50\%
\\
\hline
\end{tabular} 
\caption{Results on the Bitcoin blockchain for $k = 10000$ and $s = 10$.}
\label{tbl:bitcoin-k-10000-s-10}
\end{table}

\section{Discussion}
\label{sec:discussion}

\subsection{SeF Codes with Proof-of-X and Hybrid Blockchains}
\label{sec:proof-of-X}
For simplicity, we have focused our attention in this paper on proof-of-work based Nakamoto consensus that is used in Bitcoin and Ethereum. SeF codes, however, can be used with any {\it proof-of-X protocol}~\cite{Bano:consensus:17}, such as proof-of-stake~\cite{Ouroboros:17} or proof-of-space~\cite{Proof-of-space:13}, with minimal changes. Essentially, a proof-of-X protocol uses an energy-efficient alternative to proof-of-work to build a chain based on the longest chain rule, similar to Bitcoin and Ethereum. SeF codes can be used with any such protocol that allows a node to verify the validity of consensus rules for each block individually. For instance, a node should be able to verify that the block creator has spent a certain amount of a resource uniquely for the block. 

In contrast to protocols that grow their chains based on the longest chain rule allowing forks, a class of protocols that avoids forks are called hybrid blockchain protocols, see, \eg~\cite{Hybrid:17,Solioda:17,OmniLedger:18,RapidChain:18,Snow-white:16,Ouroboros:17,Algorand:17}. Such a protocol typically elects a {\it committee} of block validators and relies on classical Byzantine fault-tolerant (BFT) consensus protocols (\eg,~\cite{PBFT:99}). These committees are usually re-elected at a slower rate than the rate at which transaction blocks are added to the blockchain. The protocol also creates a special type of blocks, called {\it identity blocks}, that contains the list of committee members. Specifically, every identity block contains the list of members of a new committee, signed by the previous committee. When SeF codes are used with a hybrid protocol, a new node will first need to download and verify every identity block before error-resilient peeling decoding can be performed.

\begin{figure}
    \centering
    \includegraphics[scale=0.5]{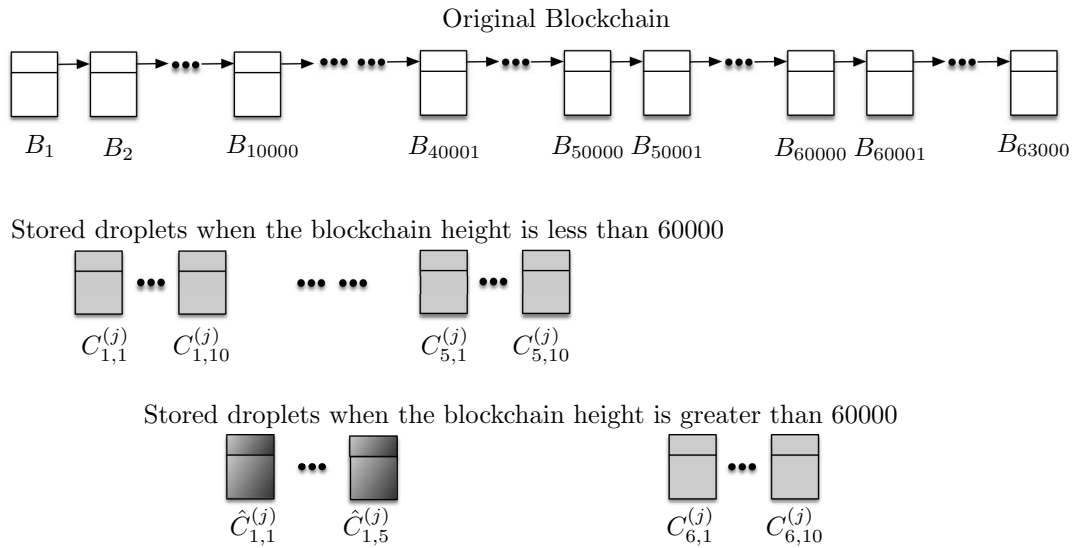}
    \caption{{\it Using multiple increasing epoch lengths to achieve dynamic storage savings. As an example, we consider $(k_1 = 10000, s_1 = 10), (k_2 = 50000, s_1 = 5)$. In every small epoch of length $k_1$, droplet nodes compute droplets using a SeF code with parameters $(k_1 = 10000, s_1 = 10)$. After a period of five small epochs (which we call a long epoch), a droplet node acts as a new node, collects droplets for each of the five previous small epochs, and decodes the blockchain for these epochs. Then, it re-encodes the blockchain using a SeF code with parameters $(k_2 = 50000, s_2 = 5)$, and deletes the droplets corresponding to the small epochs.}}
    \label{fig:encoding-2-tier}
\end{figure}

\subsection{Achieving Dynamic Storage Savings}
\label{sec:higher-storage-savings}
One limitation of our current proposal is that SeF codes are tuned to achieve a predetermined storage savings by fixing the epoch length $k$ and the number of droplets stored per epoch $s$. An easy way to achieve dynamic storage savings is to allow droplet nodes to choose any $s \geq 1$ depending on their storage budget. In this way, a node can achieve any storage saving between $\{k, k/2, k/3, \ldots, 1\}$. In fact, a node can choose different values of $s$ for different epochs. One natural way is to choose a large $s$ (\eg, $s = 10$) for all epochs, and then decrease $s$ for older epochs by deleting randomly selected droplets in those epochs. 

Additionally, it is possible to choose multiple pairs $(k_i, s_i)$ with increasing epoch lengths, and perform encoding for longer epochs in the background. To be specific, let us consider an example of $(k_1 = 10000, s_1 = 10), (k_2 = 50000, s_2 = 5)$. A droplet node encodes small epochs using SeF codes with parameters $(k_1 = 10000, s_1 = 10)$. After a period of five small epochs, \ie, when the blockchain grows by $k_2$ (which we call as a long epoch), it acts as a new node, collects droplets for each of the five previous small epochs, and decodes the blockchain for these epochs. Then, it re-encodes the decoded blockchain using a SeF code with parameters $(k_2 = 50000, s_2 = 5)$, and deletes the droplets corresponding to the small epochs. (See Fig.~\ref{fig:encoding-2-tier}.)
A bucket node joining the network downloads droplets for older long epochs and recent small epochs. For instance, consider a new node joining the network when the height of the longest chain is $t = 63000$. Then, a bucket node collects droplets corresponding to a SeF code with $(k_2 = 50000, s_2 = 5)$ for the first long epoch, and droplets corresponding to a SeF code with $(k_1 = 10000, s_1 = 10)$ for the sixth smaller epoch. (See Fig.~\ref{fig:encoding-2-tier}.)
Note here that, by decoding and re-encoding for longer epochs in the background, droplet nodes are trading-off computation as well as communication for increasing their storage savings.

\subsection{Reducing Bandwidth Overhead by Downloading Droplets As Needed}
\label{sec:as-needed}
It is possible to significantly reduce the bandwidth overhead by selectively downloading droplets. This is especially easy in the case of random sampling. Specifically, after contacting a droplet node, a bucket node can first query just the indices of the droplets that it is storing. Then, it will download only the droplets that it has not previously downloaded. This allows a bucket node to reduce its bandwidth overhead close to the minimum (assuming that the queries occupy relatively small bandwidth compared to the block-size).

Similar idea can be used to reduce the bandwidth overhead for SeF codes. 
In particular, a bucket node will first download only the binary vectors $\vj{j}$'s from a large number of droplet nodes. Then, it starts decoding by forming a bipartite graph $G$ using the binary vectors (see Step 1). In every iteration, if there exists a droplet that will result in a singleton, it downloads that particular droplet by contacting the droplet node which provided the corresponding binary vector. Here we assume that it is possible to re-contact droplet nodes. We list the bandwidth overhead incurred by this algorithm in  Table~\ref{tbl:bitcoin-as-needed}. 

\begin{table}[t!]
\centering
\begin{tabular}{|c|c|c|c|c|}
\hline
{} & 
\begin{tabular}{@{}c@{}} 
Adaptive Zero  \\ Padding
\end{tabular} 
& 
\begin{tabular}{@{}c@{}} 
Block Concatenation \\ to 1MB
\end{tabular} 
& 
\begin{tabular}{@{}c@{}} 
Block Concatenation \\ to 5MB
\end{tabular} 
&
\begin{tabular}{@{}c@{}} 
Block Concatenation \\ to 10MB
\end{tabular} 
\\
\hline
\hline
$k=1000$, $s=1$
& 
31.92\%
& 
11.16\%
& 
3.88\%
&
2.07\%
\\
\hline
$k=10000$, $s=10$
& 
33.19\%
& 
11.45\%
& 
4.24\%
&
2.35\%
\\
\hline
\end{tabular} 
\caption{Average bandwidth overhead for SeF codes on the Bitcoin blockchain when downloading droplets ``as needed'' . We consider the case that all droplet nodes are honest.}
\label{tbl:bitcoin-as-needed}
\end{table}

\subsection{Dealing with Non-Oblivious Adversary}
\label{sec:non-oblivious}
As we showed in Sec.~\ref{sec:analysis-SeF}, SeF codes are secure against an oblivious adversary that does not observe storage contents of droplet nodes before choosing which nodes to control. However, a non-oblivious adversary can corrupt a limited number of nodes to induce decoding failure for SeF codes. As an example, consider the following {\it bribery attack}. An adversary first acts as a bucket node to learn about the storage of a large number of honest droplet nodes. Then, it uses this information to corrupt (bribe) a subset of nodes. Such an adversary can induce decoding failure by, for example, bribing droplet nodes that store at least one singleton droplet. In this case, it is easy to see that the adversary needs to bribe only $O(\sqrt{k}\ln\left(\frac{k}{\delta}\right))$ droplet nodes out of $k + O\left(\sqrt{k}\ln^2(k/\delta)\right)$ ones to induce decoding failure (assuming $s = 1$ for simplicity). This is because $k + O\left(\sqrt{k}\ln^2(k/\delta)\right)$ clear droplets contain $O(\sqrt{k}\ln\left(\frac{k}{\delta}\right))$ singleton droplets on average (see~\eqref{eq:robust-soliton}). We leave the problem of designing computationally efficient coding schemes that are secure against a non-oblivious adversary as a future work.

It is worth noting that, in a typical blockchain network, new nodes will keep joining the network. If new honest nodes join the network at a rate that is greater than the rate at which adversary can observe and control nodes, then SeF codes will be secure.

\section*{Acknowledgement}
\label{sec:ack}
S. Kadhe would like to thank  O. Ozan Koyluoglu for helpful comments on initial drafts of this paper.

\bibliographystyle{IEEEtran}
\bibliography{Bib_codChains,IEEEabrv}

\appendix

\section{Proof of Lemma~\ref{lem:decoding-condition}}
\label{app:proof-of-lemma}

The proof relies on three propositions. The first two propositions establish the behavior of the decoder in Step (3). Note that in every iteration, the error-resilient peeling decoder either decodes a block or a deletes a droplet in Step (3). First, we show that it never incorrectly decodes a block. Next, we show that it never deletes a droplet that is not murky. For simplicity, we consider the first epoch.

\begin{proposition}
\label{prop:correct-decoding}
If the error-resilient peeling decoder decodes a block in Step (3), it must be a correct block. 
\end{proposition}
\begin{proof}
Consider an iteration in which the decoder decodes the $m$-th block to $\hblock{m}$. Let $\cblockj{l} = \{\header{l},\payload{l}\}$ be the singleton droplet connected to $\block{m}$ in $G^{i-1}$ in Step (2). Thus, we have $\hblock{m} = \cblockj{l}$. 

Suppose, for contradiction, that $\hblock{m} \ne \block{m}$, which, in turn, gives $\cblockj{l} \ne \block{m}$. Now, from Step (3a), we must have $\header{l} = \header{m}$, and $\mroot{\payload{l}} = \mroot{\payload{m}}$. Since every droplet node is assumed to be computationally bounded and $\hash{\cdot}$ is cryptographically secure, we must have $\payload{l} = \payload{m}$. Thus, $\cblockj{l} = \block{m}$, resulting in a contradiction.
\end{proof}

\begin{proposition}
\label{prop:deleting-droplet}
If the error-resilient peeling decoder rejects a droplet in Step (3), then it must a murky droplet.
\end{proposition}
\begin{proof}
Consider an iteration $i$ in which the decoder deletes a droplet $\cblockj{l} = \{\header{l},\payload{l}\}$. Suppose that the decoder has decoded $s-1$ blocks until that iteration, for some $1\leq s \leq k$. Denote the decoded blocks as $\hblock{j_1},\hblock{j_2},\ldots,\hblock{j_{s-1}}$.
From Step (2), $\cblockj{l}$ must be connected to exactly one block in $G^{i-1}$. Denote that block as $\block{j_s}$. Let $d$ be the degree of $\cblockj{l}$ in $G$ at the initialization Step (1). 

{\it Case 1:} $d = 1$. The length-$k$ vector $\vj{l}$ associated with $\cblockj{l}$ is such that its $j_s$-th entry is $1$ and every other entry is $0$. Suppose, for contradiction, that $\cblockj{l}$ is not murky. That is, $\cblockj{l} = \vj{l}\mathbf{B} = \block{j_s}$. However, since the decoder deletes $\cblockj{l}$, either $\header{l} \ne \header{j_s}$ or $\mroot{\payload{l}} \ne \mroot{\payload{j_s}}$ (or both), which results in a contradiction. 
Hence, $\cblockj{l}$ must be murky.

{\it Case 2:} $d\geq 2$. Since, at iteration $i$, $\cblockj{l}$ is connected to only $\block{j_s}$, it must be that $d\leq s$ and the other $d-1$ neighbors of $\cblockj{l}$ form a subset of $\block{j_1},\block{j_2},\ldots,\block{j_{s-1}}$. Without loss of generality, let $\block{j_1},\block{j_2},\ldots,\block{j_{d-1}},\block{j_s}$ be the $d$ neighbors of $\cblockj{l}$ in $G$ at the initialization. In other words, the length-$k$ vector $\vj{l}$ is such that its $i$-th entry is $1$ for $i = j_1, j_2,\ldots, j_{d-1}, j_s$, and every other entry is $0$.
Moreover, at iteration $i$, we have $\cblockj{l} = \cblockj{l}^{(0)}\oplus\hblock{j_1}\oplus\hblock{j_2}\oplus\cdots\oplus\hblock{j_{d-1}}$, where $\cblockj{l}^{(0)}$ be the value of the droplet at the initialization. By Proposition~\ref{prop:correct-decoding}, each of the $s-1$ decoded blocks are correct, and thus, $\cblockj{l} = \cblockj{l}^{(0)}\oplus\block{j_1}\oplus\block{j_2}\oplus\cdots\oplus\block{j_{d-1}}$.

Suppose, for contradiction, that $\cblockj{l}$ is not murky. That is, at the beginning of the decoding, we have $\cblockj{l}^{(0)} = \vj{l}\mathbf{B} = \block{j_1}\oplus\block{j_2}\oplus\cdots\oplus\block{j_{d-1}}\oplus\block{j_s}$. Thus, at iteration $i$, we must have $\cblockj{l} = \block{j_s}$. However, since the decoder deletes $\cblockj{l}$, either $\header{l} \ne \header{j_s}$ or $\mroot{\payload{l}} \ne \mroot{\payload{j_s}}$ (or both), which results in a contradiction. 
Hence, $\cblockj{l}$ must be murky.
\end{proof}

Recall that Step (3) differentiates the error-resilient peeling decoder from the classical peeling decoder for an LT code~\cite{Luby:02}.
In particular, in contrast to the classical peeling decoder which always {\it accepts} a singleton, the error-resilient peeling decoder may {\it reject} a singleton if its header and/or Merkle root does not match with the one stored in the header-chain. 
Now, suppose that we could identify the subset of clear droplets $\tS$ among the set of collected droplets $S$ at the beginning of the decoding. Then, we can use the classical peeling decoder to recover the blockchain from these clear droplets $\tS$. In the following proposition, we show that if the classical peeling decoder succeeds to recover the entire blockchain from $\tS$, then the error-resilient peeling decoder must succeed on $S$, even though it is not possible for the decoder to identify the clear droplets at the beginning of the decoding. 

\begin{proposition}
\label{prop:success-on-clear-droplets}
Let $S$ denote a set of droplets corresponding to an arbitrary epoch that are collected by a bucket node, and $\tS$ denote the subset of clear droplets from $S$. If the classical peeling decoder can recover the blockchain for the epoch from $\tS$, then the error-resilient decoder must be able to recover the blockchain for the epoch from $S$.
\end{proposition}
\begin{proof}
First, note that the classical and error-resilient decoders are equivalent on $\tS$. This is because the error-resilient peeling decoder will never delete a droplet from $\tS$, since all the droplets are clear (see Proposition~\ref{prop:deleting-droplet}). Therefore, it suffices to focus only on the error-resilient decoder in the proof. In other words, it suffices to show that  if the error-resilient peeling decoder succeeds to decode the epoch from $\tS$, it will also succeed to decode the epoch from $S$.

Note that any block decoded from $S$ must be correct by Proposition~\ref{prop:correct-decoding}. Thus, it is sufficient to show that if the error-resilient peeling decoder does not declare failure when decoding from $\tS$, it will not declare failure when decoding from $S$. 





Let $G$ and $\tilde{G}$ be the bipartite graphs in Step (1) when decoding from $S$ and $\tS$, respectively. Now, since decoding with $\tilde{G}$ as the starting point succeeds, at each iteration $i$, $1\leq i\leq k$, there is at least one singleton droplet in $\tilde{G}^{i-1}$. Note that this happens irrespective of which singleton was chosen in the previous iteration, because, if there are multiple singletons available in an iteration, the choice of the singleton does not affect the success of the decoder in recovering the blockchain.

Availability of at least one singleton droplet while decoding from $\tilde{G}$ implies that, when decoding with $G$ as the starting point, there must be at least one clear singleton droplet on $G^{j-1}$ at every iteration $j$. This is because deleting a murky droplet does not change the degree of any clear singleton, and accepting a singleton corresponding to an opaque droplet can only reduce the degree of some clear droplets, which in turn helps in creating clear singletons. Therefore, if the error-resilient peeling decoder does not declare failure when decoding from $\tS$, it will not declare failure when decoding from $S$. This completes the proof.
\end{proof}


Now, we are ready to prove Lemma~\ref{lem:decoding-condition}. 
First, note that the bucket node has at least $k + O\left(\sqrt{k}\ln^{2}(k/\delta)\right)$ clear droplets, as the set of droplet nodes it contacts contains at least $\frac{1}{s}\left(k + O\left(\sqrt{k}\ln^{2}(k/\delta)\right)\right)$ honest nodes. Let us denote the set of clear droplets as $\tS$.
Further, note that the adversary cannot influence the probability of decoding failure from $\tS$. This is because the adversary corrupts droplet nodes without observing their storage contents, and thus, it is oblivious to the contents of the honest nodes.
Now, from~\cite[Theorem 17]{Luby:02}, it follows that the probability that the classical peeling decoder fails to recover the $k$ blocks of an arbitrary epoch from $\tS$ is at most $\delta$. Now, recall that we assume that the same randomness is used for encoding every epoch. Thus, the recovery of an arbitrary epoch ensures the recovery of all the epochs. Therefore, the classical peeling decoder will fail to recover the blockchain from $\tS$ with probability at most $\delta$. Finally, using Proposition~\ref{prop:success-on-clear-droplets} completes the proof of Lemma~\ref{lem:decoding-condition}.

\section{Proof of Theorem~\ref{thm:SeF-performance}}
\label{app:proof-of-theorem-1}
Decentralization follows directly from the property of LT codes that the degree and neighbors for every droplet are chosen independent of the other droplets. Therefore, a droplet node does not need to rely on any other node in the network while computing its droplets.

It is easy to see that the storage savings is $k/s$: each droplet node stores $s$ droplets whenever the blockchain grows by $k$ blocks. Here we use the assumption that all blocks are of the same size together with assumption (i).

The bootstrap cost immediately follows from Lemma~\ref{lem:decoding-condition}. 

To prove the bandwidth overhead, it is sufficient to show that it possible to recover the blockchain with high probability by contacting $n = \frac{K(k/s,\delta)}{1-\sigma}$ droplet nodes. Towards this end, let $\epsilon = \sqrt{\frac{2\ln{\delta}}{(1-\sigma)n}}$. Now, assumption (ii) states that the probability that each of the contacted droplet node is honest is $(1-\sigma)$ independent of the others. Thus, using the Chernoff bound, the probability that these $n$ nodes contain smaller than $(1-\epsilon)(1-\sigma)n$ honest nodes is at most $e^{-\epsilon^2(1-\sigma)n/2}$. Combining this with Lemma~\ref{lem:decoding-condition}, it is not hard to show that the probability of successfully decoding the blockchain from the $ns$ droplets is at least $1 - 2\delta$. 

Finally, the computation cost follows from the properties of the LT codes as shown in~\cite{Luby:02}. In particular, it shown in~\cite[Theorem 13]{Luby:02} that the average degree of a droplet is $O(\ln(k/\delta))$. Thus, it takes $O(s\ln(k/\delta))$ operations on average to compute $s$ droplets. This give the encoding cost. To compute the decoding cost, note that it is proportional to the average number of edges in the graph $G$ formed at the beginning of decoding. (Recall assumption (iii) that we do not consider the cost of computing Merkle roots.) The average number of edges can be easily computed by noting that the average number of droplets sufficient to recover the blockchain with high probability is $\frac{sK(k/s,\delta)}{1-\sigma}$, and each droplet is of degree $O(\ln(k/\delta))$ on average.

\newpage
\section{Details of Experimental Results on the Bitcoin Blockchain}
\label{app:experiment-details}

\begin{table}[h!]
\centering
\begin{tabular}{c||c}
Super-block size                & No concatenation\\
Number of blocks              & 565876\\
Number of epochs               &   565\\
Original blockchain size       & 197063.58MB\\
Average storage per node  &       262.95MB\\
Average download size $(\sigma = 0)$ & 296748.18MB\\
Average download size $(\sigma = 0.1)$ & 329693.39MB\\
\hline
Super-block size                & 1MB\\
Number of super-blocks     & 220254\\
Number of epochs              &   220\\
Original blockchain size       & 197677.34MB\\
Average storage per node  &       220.61MB\\
Average download size $(\sigma = 0)$ & 249012.71MB\\
Average download size $(\sigma = 0.1)$ & 276640.20MB\\
\hline
Super-block size                & 5MB\\
Number of super-blocks     & 42843\\
Number of epochs              &   42\\
Original blockchain size       & 194142.21MB\\
Average storage per node  &       201.95MB\\
Average download size $(\sigma = 0)$ & 227821.25MB\\
Average download size $(\sigma = 0.1)$ &  253057.85MB\\
\hline
Super-block size                & 10MB\\
Number of super-blocks     & 20688\\
Number of epochs              &   20\\
Original blockchain size       & 191480.81MB\\
Average storage per node  &       195.60MB\\
Average download size $(\sigma = 0)$ &  220816.31MB\\
Average download size $(\sigma = 0.1)$ &  245039.65MB\\
\end{tabular} 
\caption{Simulations on the Bitcoin blockchain for $k = 1000$ and $s = 1$. (The number of epochs denote the number of past epochs. The current epoch is excluded while computing the original blockchain size and the average download size.)}
\label{tbl:bitcoin-k-1000-s-1-details}
\end{table}

\begin{table}[h!]
\centering
\begin{tabular}{c||c}
Super-block size                & No concatenation\\
Number of blocks              & 565876\\
Number of epochs               &   56\\
Original blockchain size       & 192105.30MB\\
Average storage per node  &       257.93MB\\
Average download size $(\sigma = 0)$ & 270278.67MB\\
Average download size $(\sigma = 0.1)$ & 300407.17MB\\
\hline
Super-block size                & 1MB\\
Number of super-blocks     & 220254\\
Number of epochs              &   22\\
Original blockchain size       & 197677.34MB\\
Average storage per node  &       221.00MB\\
Average download size $(\sigma = 0)$ & 231485.75MB\\
Average download size $(\sigma = 0.1)$ & 257355.23MB\\
\hline
Super-block size                & 5MB\\
Number of super-blocks     & 42843\\
Number of epochs              &   4\\
Original blockchain size       & 185091.69MB\\
Average storage per node  &       193.09MB\\
Average download size $(\sigma = 0)$ & 202225.28MB\\
Average download size $(\sigma = 0.1)$ &  225054.46MB\\
\hline
Super-block size                & 10MB\\
Number of super-blocks     & 20688\\
Number of epochs              &   2\\
Original blockchain size       & 191480.81MB\\
Average storage per node  &       196.07MB\\
Average download size $(\sigma = 0)$ & 205516.55MB\\
Average download size $(\sigma = 0.1)$ &  228827.68MB\\
\end{tabular} 
\caption{Simulations on the Bitcoin blockchain for $k = 10000$ and $s = 10$. (The number of epochs denote the number of past epochs. The current epoch is excluded while computing the original blockchain size and the average download size.)}
\label{tbl:bitcoin-k-1000-s-1-details}
\end{table}

\end{document}